\documentclass[12pt,preprint]{aastex}
\received{}
\revised{}
\accepted{}
\newcommand{\beq}{\begin{equation}}
\newcommand{\eeq}{\end{equation}}
\newcommand{\ba}{\begin{array}}
\newcommand{\ea}{\end{array}}

\newcommand{\cm}{\mbox{ cm}}
\newcommand{\eV}{\mbox{ eV }}
\newcommand{\keV}{\mbox{ keV }}
\newcommand{\MeV}{\mbox{ MeV }}
\newcommand{\GeV}{\mbox{ GeV }}

\begin{document}
\title {Prompt GRB spectra: detailed calculations and the effect of pair production}
\author{Asaf Pe'er\altaffilmark{1}\altaffilmark{2} and Eli Waxman\altaffilmark{1}}
\altaffiltext{1}{Physics Faculty, Weizmann
Institute, Rehovot 76100, Israel}
\altaffiltext{2}{asaf@wicc.weizmann.ac.il}

% --------------------------------------------------------------------------
% Version Notes:
% First draft, July 10th, 2003. 
% --------------------------------------------------------------------------

\begin{abstract}

We present detailed calculations of the prompt spectrum of gamma-ray bursts (GRBs) 
predicted within the fireball model framework,
where emission is due to internal shocks in an expanding relativistic
wind. Our time dependent numerical model describes cyclo-synchrotron
emission and absorption, inverse and direct Compton scattering, and
$e^\pm$ pair production and annihilation (including the evolution of
high energy electro-magnetic cascades). It allows, in particular, a
self-consistent calculation of the energy distribution of $e^\pm$
pairs produced by photon annihilation, and hence a calculation of the
spectra resulting when the scattering optical depth due to pairs, $\tau_\pm$, is
high. We show that emission peaks at $\sim1$~MeV for moderate to large $\tau_\pm$, reaching  $\tau_\pm\sim 10^2$. In this regime of large compactness we find that 
(i) A large fraction of shock energy can escape as radiation even for large $\tau_\pm$; 
(ii) The spectrum depends only weakly on the magnetic field energy fraction;
(iii) The spectrum is hard, $\varepsilon^2 dN/d\varepsilon\propto \varepsilon^\alpha$ 
with $0.5<\alpha<1$, between the self absorption ($\varepsilon_{ssa}= 10^{0.5\pm0.5}\keV$) 
and peak ($\varepsilon_{peak}= 10^{0.5\pm0.5} \MeV$) photon energy, 
(iv) and shows a sharp cutoff at $\sim10$~MeV; 
(v) Thermal Comptonization leads to emission peaking at $\varepsilon_{peak} \gtrsim30$~MeV, 
and can not therefore account for observed GRB spectra.
For small compactness, spectra extend to $>10$~GeV with flux detectable by GLAST, and the spectrum at low energy depends on the magnetic field energy fraction. 
Comparison of the flux at $\sim1$~GeV and $\sim100$~keV may therefore allow 
to determine the magnetic field strength.
For both small and large compactness, the spectra depend only weakly on the spectral index 
of the energy distribution of accelerated electrons. 

\end{abstract}

% --------------------------------------------------------------------------
% Section: Introduction
% --------------------------------------------------------------------------

\section{Introduction}
\label{sec:intro}
\indent

It is widely accepted that $\gamma$-ray bursts are produced by the
dissipation of kinetic energy in a highly relativistic wind, driven by
gravitational collapse of a (few) solar mass object into a neutron
star or a black hole (see
e.g. \citet{fireballs1,fireballs2,fireballs3} for reviews). The prompt
$\gamma$-ray emission is believed to be produced by synchrotron and
inverse-Compton emission of electrons accelerated to relativistic
energy by internal shocks within the expanding wind
\citep[see, however,][]{LGCR00,GLCR00}. Synchrotron emission 
is favored if the fireball is required to be "radiatively efficient,"
i.e. if a significant fraction of the fireball energy is required to
be converted to $\gamma$-rays. 

Over a wide range of model parameters, a large number of $e^\pm$ pairs
are produced in internal collisions, due to annihilation of high
energy photons \citep[e.g.][]{MR00, GSW01}. In fact, if the
internal shocks occur at small enough radii, the plasma becomes
optically thick and a second photosphere is formed \citep{MR00,MRRZ}
beyond the photosphere associated with the electrons initially present
in the fireball. 
As we show here \citep[\S~\ref{sec:compactness}, see also ][]{GSW01}
requiring the emission to be dominated by $\sim1$~MeV photons implies,
within the fireball model framework, a moderate to large optical depth
due to scattering by pairs.
 When the scattering optical depth due to pairs is
high, calculation of the emergent spectrum becomes
complicated. Relativistic pairs cool rapidly to mildly-relativistic
energy, where their energy distribution is determined by a balance
between emission and absorption of radiation. The emergent spectrum,
which is affected by scattering off the pairs population, depends
strongly on the pairs energy distribution, and in particular on the
"effective temperature" which characterizes the low-end of the energy
distribution. The pairs energy distribution is difficult to calculate
analytically. Moreover, analytic calculation of the spectrum emerging
from the electromagnetic cascades initiated by photon annihilation is
also difficult. Therefore, in order to derive the emergent spectrum a
numerical model is needed \citep[see e.g.][]{Ghisellini99, Zhang03}.   

Emission from steady plasma, 
where pair production and annihilation are taken into consideration, 
was numerically studied in
the past in the context of active galactic nuclei (AGN's). 
It was found that thermal plasma, 
optically thin to Thomson scattering,
and characterized by a comoving compactness 
$10 \lesssim l' \lesssim 10^3$,
has a normalized pair temperature 
$\theta \equiv kT/m_ec^2 \approx 10^{-2} - 10^{-1}$ 
\citep{Lightman82, Svensson82, Svensson84}. 
The dimensionless compactness parameter $l$ is defined by 
 $l \equiv L \sigma_T / R m_e c^3$, 
where  $L$ is the luminosity, 
and $R$ is a characteristic length of the object.  
The above result holds also in a scenario considering injection of high
energy particles, 
which lose their energy via IC scattering of low energy photons, and
thermalize before annihilating \citep{Svensson87, LZ87}.
The optical depth for scattering by pairs was found in the above analyses to be 
$0.1 \lesssim \tau_\pm \lesssim 15$. However, $\tau_\pm$ 
strongly depends on the comoving compactness $l'$, 
and sharply increases beyond these values 
when the compactness increases beyond $10^3$ \citep[see][]{ LZ87, Svensson87}. 
When the scattering optical depth $\tau_\pm \gg 1$ and $\theta \ll 1$, solving the Kompaneets 
equation gives a cutoff at normalized photon energy $h\nu/m_ec^2 
\simeq 1/\tau_\pm^2$ \citep{Sunyaev80}.

The results mentioned above are not directly applicable to GRB
plasma. The GRB plasma is rapidly 
evolving and steady state can not be assumed. For example: The
electron distribution function does not reach thermal equilibrium even
at low energy (see \S~\ref{sec:results}); Due to expansion, the
average number of scattering a photon undergoes is $\sim\tau_\pm$ rather 
than $\tau_\pm^2$; The expansion (and cooling) of plasma electrons has
a significant effect on the photon spectrum when $\tau_\pm$ is high.
Moreover, significant luminosity at high energies (exceeding the pair production threshold)
 is expected due to synchrotron emission from energetic particles in strong magnetic fields,
 $B\sim 10^5 - 10^6$~G, typical for GRB shell-shell collision phase, 
a phenomenon that was not considered in the analyses mentioned above. 
And last, in the GRB case a non-thermal high energy electron population is assumed 
to be produced, which leads at large compactness to the formation of a
rapid electro-magnetic cascade, the evolution of which was not
considered in the past.

A numeric calculation of GRB spectra that takes into consideration creation 
and annihilation of pairs is complicated. 
The evolution of electromagnetic cascades initiated by
the annihilation of high energy photons occurs on a very short time scale.
On the other extreme, evolution of the low-energy,
mildly relativistic pairs, which is governed by synchrotron
self-absorption, direct and inverse Compton emission takes much
longer time. 
The large
difference in characteristic time scales poses a challenge to numeric
calculations. For this reason, the only numerical approach so far
\citep{Pilla98} 
was based on using a Monte-Carlo method. 
This model, however, did  not consider the
parameter space region where pairs strongly affect the spectra. 
Another challenge to numerical modeling is due to the fact that at
mildly relativistic energies the usual synchrotron emission and IC
scattering approximations are not valid, and a precise
cyclo-synchrotron emission and direct- inverse Compton scattering 
calculations are required. 

In this work, we consider emission within the fireball model
framework, resulting from internal shocks 
within an expanding relativistic wind.
These shock waves dissipate kinetic energy,
 and accelerate a population of relativistic electrons. 
We adopt the common assumption of a power law energy
distribution of the accelerated particles, and calculate the
emergent spectra.
We present the results of time dependent
numerical calculations, considering all the relevant physical
processes: cyclo-synchrotron emission, synchrotron self absorption,
inverse and direct Compton scattering, and $e^\pm$ pair production and
annihilation including the evolution of high energy electro-magnetic
cascades. A full description of the numerical code appears in
\citet{Pe'er03b}. 
   
We note, that Comptonization by a thermal population of electrons (and possibly $e^\pm$ pairs)
was considered as a possible mechanism for GRB production
\citep[e.g. ][]{Ghisellini99}, following the  
evidence that, at least in some cases, the GRB spectra at low energy
is steeper 
at early times than expected for synchrotron emission
\citep{Crider,Preece,Frontera}. 
This model is different than the common model considered in the
previous paragraph in assuming that the kinetic energy dissipated in 
a collision between two "shells" within the expanding wind 
is continuously distributed among all shell electrons,
rather than being deposited at any given time into a small fraction 
of shell electrons that pass through the shock wave, 
and in assuming that energy is equally distributed among electrons,
rather than following a power-law distribution. 
There is no known acceleration mechanism that leads to the above energy  
distribution among electrons, and the spectrum predicted in this scenario does not account for the claimed steep spectra,
which may be naturally explained as a contribution to the observed
$\gamma$-ray radiation of photospheric fireball emission
\citep[e.g. ][]{MR00}. It is nevertheless worthwhile to derive the spectrum that is
expected from thermal Comptonization under plasma conditions typical
to GRB fireballs. This will allow to determine whether this process
may significantly contribute to GRB $\gamma$-ray emission.  
Our numerical code enables an accurate calculation of the 
pair temperature in this scenario, as well as reliable
calculation of the emergent spectrum.  

This paper is organized as follows. In \S\ref{sec:parameters} we derive the plasma parameters at the internal shocks stage, and their dependence on uncertain model parameters values. 
In \S~\ref{sec:compactness} it is shown that moderate to large compactness is expected for the parameter range where emission peaks at $\sim1$~MeV, and approximate analytic results describing the emission in this regime are given. 
Our numerical methods are briefly presented in \S\ref{sec:numerical};
a detailed description of the model is found in \citet{Pe'er03b}. 
Our numerical results for the scenario of acceleration of particles in shock waves
are presented in \S\ref{sec:results}.
In \S\ref{sec:ghis} we present both analytical and numeric
calculations of the spectra resulting from thermal Comptonization. 
We summarize and conclude our discussion in \S\ref{sec:summary},
with special emphasis on observational implications.

\section{Plasma parameters and large compactness behavior}
\label{sec:analytic}

Variability in the Lorentz factor of the relativistic wind emitted by the GRB progenitor leads to the formation of shock waves within the expanding wind, at large radii compared to the underlying source size. Denoting by $\Gamma$ the characteristic wind Lorentz factor and assuming variations $\Delta\Gamma/\Gamma\sim1$ on time scale $\Delta t$, shocks develop at radius $r_i\approx2\Gamma^2 c \Delta t = 5.4 \times 10^{11} \ \Gamma_{2.5}^2 \  \Delta t_{-4} \cm$. For $\Delta\Gamma/\Gamma\sim1$ the shocks are mildly relativistic in the wind frame. For our calculations, we consider a collision between two uniform shells of thickness $c\Delta t$, in which two mildly relativistic ($\Gamma_s-1\sim1$ in the wind frame) shocks are formed, one propagating forward into the slower shell ahead, and one propagating backward (in the wind frame) into the faster shell behind. 
The comoving shell width, measured in the shell rest frame, 
is $\Delta R=\Gamma c \Delta t$, and the comoving dynamical time, 
the characteristic time for shock crossing and shell expansion measure 
in the shell rest frame, is $t_{dyn}=\Gamma\Delta t$.

Under these assumptions, the shocked plasma conditions are determined by 6 model
parameters. Three are related to the underlying source:
The total luminosity $ L = 10^{52}\,L_{52} \rm{\ erg \,s^{-1}}$, 
the Lorentz factor of the shocked plasma
$\Gamma = 10^{2.5}\,\Gamma_{2.5}$, 
and the variability time $\Delta t = 10^{-4}\, \Delta t_{-4} \rm{\,s}$. 
Three additional parameters are related to the collisionless-shock microphysics:
The fraction of post shock thermal energy carried by electrons
$\epsilon_e = 10^{-0.5}\, \epsilon_{e,-0.5}$ and by magnetic field $\epsilon_B = 10^{-0.5}\, \epsilon_{B,-0.5}$,
and the power law index of the accelerated electrons energy distribution, 
$d\log n_e/d\log\varepsilon_e=-p$. 
In the following calculations spherical geometry is assumed. 
However, the results are valid also for a jet like GRB, provided
the jet opening angle $\theta > \Gamma^{-1}$ (in which case $L$ should be regarded
as the isotropic equivalent luminosity).

In \S~\ref{sec:parameters} below we derive the characteristic values of plasma parameters obtained under the adopted model assumptions, and the characteristic synchrotron and self-absorption frequencies.
Since, as we show in \S~\ref{sec:compactness}, the compactness parameter is related to the optical depths for both pair production and scattering by pairs, emission from plasma characterized by small compactness is well approximated by the optically thin synchrotron-self-Compton emission model. This model, however, ceases to be valid for large value of the compactness. In \S~\ref{sec:compactness} we give an approximate analysis of the emission of radiation from moderate to large compactness plasma. 
 
\subsection{Plasma parameters}
\label{sec:parameters}

Denoting by $\theta_{p}$ the average proton internal energy (associated with random motion) in the shocked plasma, measured in units of the proton's rest mass, the comoving proton density in the shocked plasma is given by
\beq
n_p \approx {L \over 4 \pi r_i^2 c \Gamma^2 \theta_{p} m_p c^2} = 
6.7 \times 10^{14} \ L_{52} \  \Gamma_{2.5}^{-6} \ 
 \Delta t_{-4}^{-2} \theta_{p}^{-1} \  \cm^{-3}.
\eeq
$\theta_{p}\sim1$ for mildly relativistic shocks and is limited within the fireball model framework to $\theta_{p}\lesssim$~few, since the Lorentz factors of the internal shocks can not be larger than a few. This is due to the fact that the Lorentz factors of colliding shells can not differ by significantly more than an order of magnitude: Shells' Lorentz factors are limited to the range of $\sim100$ to few thousands, where the lower limit of $\sim100$ is set by the requirement to avoid too large optical depth, and the upper limit of few~$\times10^3$ is due to the fact that shells can not be accelerated by the radiation pressure to Lorentz factors $\gg10^3$ \citep[e.g. \S2.3 in][]{fireballs3}.

A fraction $\epsilon_B$ of the internal energy density
$u_{int} = L/(4\pi r_i^2 c \Gamma^2)$
is assumed to be carried by the magnetic field, implying
\beq
B = \sqrt{\frac{\epsilon_B L}{2 \Gamma^6 c^3 \Delta t ^2}} = 
2.9\times 10^6 \ L_{52}^{1/2} \  \epsilon_{B,-0.5}^{1/2} \ 
 \Gamma_{2.5}^{-3} \  \Delta t_{-4}^{-1} \  \rm{G}.
\label{eq:B}
\eeq

Equating the particle acceleration time,  
$t_{acc}\simeq \varepsilon/(c q B)$ and the synchrotron cooling time 
$t_{cool,sync}$ gives the maximum Lorentz factor 
of the accelerated electrons,
$\gamma_{max} = (3/2) m_e c^2 (q^3 B)^{-1/2} = 
6.9 \times 10^4 \ L_{52}^{-1/4} \  \epsilon_{B,-0.5}^{-1/4} \
 \Gamma_{2.5}^{3/2} \  \Delta t_{-4}^{1/2}$,
and the maximum observed energy of the synchrotron emitted photons,
\beq
\varepsilon_{max}^{ob.} = \hbar \frac{3}{2} \frac{q B}{m_e c}
 \gamma_{max}^2 \frac{\Gamma}{1+z} = 7 \times 10^{10} (1+z)^{-1}
\ \Gamma_{2.5}   \eV.
\eeq
The minimum Lorentz factor of the power law accelerated electrons,
given by
\beq
\gamma_{min} = \left\{ \begin{array}{ll} 
\epsilon_e \theta_{p} \left( \frac{m_p}{m_e}\right) 
{\log^{-1}\left( \frac{\varepsilon_{e,max}}{\varepsilon_{e,min}} \right)} &
 (p=2) \\
\epsilon_e \theta_{p} \left( \frac{m_p}{m_e}\right)
 \frac{p-2}{p-1} & (p>2), \end{array} \right.
\label{eq:gamma_min}
\eeq
is much larger than $\gamma_c$, the Lorentz factor of electrons that cool on 
a time scale that is equal to the dynamical time scale, which is $\gamma_c \sim 1$.  

For a typical value of 
$\log\left( \varepsilon_{e,max}/\varepsilon_{e,min} \right) \simeq 7$, 
synchrotron emission from the least energetic electrons peaks at
\beq
\varepsilon_{peak}^{ob.} =  \left\{ \begin{array}{ll}
10^5 (1+z)^{-1} \quad L_{52}^{1/2} \  \epsilon_{e,-0.5}^2 \
\epsilon_{B,-0.5}^{1/2} \
 \Gamma_{2.5}^{-2} \ \Delta t_{-4}^{-1} \ \theta_{p}^2 \ \rm{eV} & (p=2); \\
5.5 \times 10^6\left(\frac{p-2}{p-1}\right)^2  (1+z)^{-1} \quad 
L_{52}^{1/2} \  \epsilon_{e,-0.5}^2 \  \epsilon_{B,-0.5}^{1/2} \
\Gamma_{2.5}^{-2} \ \Delta t_{-4}^{-1} \ \theta_{p}^2 \ \rm{eV} & (p>2).
 \\ \end{array} \right.
\label{eq:eps_peak}
\eeq
The self absorption optical depth $\tau_\nu = \alpha_\nu \Gamma c \Delta t$,
calculated using the absorption coefficient
\beq
\alpha_\nu = \left\{ \begin{array}{ll}
1.3 \times 10^{41} \nu^{-3} \quad L_{52}^{2} \  \Gamma_{2.5}^{-12} \ 
\epsilon_{e,-0.5} \ \epsilon_{B,-0.5} \ \Delta t_{-4}^{-4}\, {\rm cm^{-1}} & (p=2); \\
1.3 \times 10^{51} \nu^{-7/2} \quad  L_{52}^{9/4} \  \Gamma_{2.5}^{-27/2} \ 
\epsilon_{e,-0.5}^2 \ \epsilon_{B,-0.5}^{5/4} \ \Delta t_{-4}^{-9/2}\, {\rm cm^{-1}} & (p=3) \\
\end{array} \right.
\eeq
\citep{Rybicki}, is smaller than 1
 at $\varepsilon_{peak}$, 
\beq
\tau_{ssa,peak} =  \left\{ \begin{array}{ll}
0.23 \quad L_{52}^{1/2} \  \epsilon_{e,-0.5}^{-5} \ \epsilon_{B,-0.5}^{-1/2} \
 \Gamma_{2.5}^{-2}  \ \theta_{p}^{-6} & (p=2); \\
8.5 \times 10^{-4} \quad 
L_{52}^{1/2} \  \epsilon_{e,-0.5}^{-5} \  \epsilon_{B,-0.5}^{-1/2} \
\Gamma_{2.5}^{-2} \ \theta_{p}^{-7} & (p=3). \\
\end{array} \right.
\eeq

If the fraction of thermal energy carried by the magnetic field is very small, 
$\epsilon_B \leq 10^{-5} \ L_{52}^{-1} \  \Gamma_{2.5}^5 \  \Delta t_{-4} \ \epsilon_{e,0.5}^{-1}$,
the electrons are in the slow cooling regime (i.e., $\gamma_{min} < \gamma_c$), 
the power radiated per unit energy below $\varepsilon_{peak}$ 
is proportional to $(\varepsilon/\varepsilon_{peak})^{1/3}$, and
the energy below which the optical depth becomes larger than 1,
$\varepsilon_{ssa} = \varepsilon_{peak} \tau_{ssa,peak}^{3/5}$ is
\beq
\varepsilon_{ssa}^{ob.} = \left\{ \begin{array}{ll}
5 \times 10^3 (1+z)^{-1} \quad 
L_{52}^{4/5} \  \epsilon_{e,-0.5}^{-1} \ \epsilon_{B,-5}^{1/5} \
 \Gamma_{2.5}^{-16/5} \ \Delta t_{-4}^{-1} \ \theta_{p}^{-8/5} \ \rm{eV} & (p=2); \\
2.5 \times 10^3  (1+z)^{-1} \quad 
L_{52}^{4/5} \  \epsilon_{e,-0.5}^{-1} \  \epsilon_{B,-5}^{1/5} \
\Gamma_{2.5}^{-16/5} \ \Delta t_{-4}^{-1} \ \theta_{p}^{-11/5} \ \rm{eV} & (p=3).
\end{array} \right.
\label{eq:eps_ssa}
\eeq 
When cooling is important ($\gamma_{min} > \gamma_c$), 
as is the case for typical fireball parameters, the electron energy distribution, and hence the 
self-absorption frequency, are modified. 
As we show below (\S~\ref{sec:results}), for large compactness the energy distribution of electrons and pairs is quasi Maxwellian.
For a thermal distribution of electrons and positrons, with normalized temperature 
$\theta \equiv kT/m_ec^2$ and normalized pair density $f\equiv n_\pm/n_p$,
the self absorption frequency $\nu_T$ is approximated by (using the results of \citet{Mahadevan96} for
cyclo-synchrotron emission) 
\beq
\nu_T = 5\times 10^{14} \quad  L_{52}^{0.6} \, \epsilon_{B,-0.5}^{0.45}
\, \Gamma_{2.5}^{-10/3} \,  \Delta t_{-4}^{-1} \,
 \theta_{-1} \, f_1^{1/6}
 \quad \rm{Hz},
\label{eq:nu_T}
\eeq
where $\theta= 0.1 \theta_{-1}$, and $f = 10 f_1$. 
This result is accurate to better than 10\%  for parameters in the range
 $0.001<L_{52}<10 ,\ 0.01<\epsilon_B\leq0.33,\ 
 100<\Gamma < 1000, 0.1<\theta<5,\  1\leq f<100$.
The values of $\theta$ and $f$ were found numerically (\S~\ref{sec:results}) to be 
 within these limits, for a wide range of parameters
that characterize GRBs.
Therefore,
\beq
\epsilon_{ssa,thermal}^{ob.} = h \nu_T \Gamma (1+z)^{-1} \approx 600 (1+z)^{-1}
\quad  L_{52}^{0.6} \, \epsilon_{B,-0.5}^{0.45}
\, \Gamma_{2.5}^{-7/3} \,  \Delta t_{-4}^{-1} \,
 \theta_{-1} \, f_1^{1/6} \quad \eV.
\label{eq:eps_ssa_thermal}
\eeq

\subsection{Large compactness behavior}
\label{sec:compactness}

The comoving compactness parameter $l'$ is defined as
$l' = \Delta R n'_\gamma \sigma_T$,
where  $\Delta R = c t_{dyn}$ is the comoving width and 
$n'_\gamma = \epsilon_e L / (4 \pi m_e c^3 \Gamma^2 r_i^2 )$ 
is the comoving number density of photons with energy exceeding the electron's rest mass, 
$\varepsilon_{ph} \geq m_e c^2$ (in the plasma rest frame). 
Only these photons are of interest, as their number density determines the number density 
of the produced pairs.
In deriving the last equation, a mean photon energy (in the comoving frame), 
$\langle\varepsilon_{ph}\rangle \approx m_e c^2$ is assumed. 
This assumption is valid as long as the spectral index $\alpha$
($\varepsilon^2 dN/d\varepsilon\propto \varepsilon^\alpha$) 
is not significantly different than 0, and leads to
\beq
\ba{ll}
l' = \frac{\epsilon_e L \sigma_T}{16 \pi m_e c^4 \Gamma^5 \Delta t} & = 
250 \ L_{52} \ \epsilon_{e,-0.5} \ \Gamma_{2.5}^{-5} \ \Delta t_{-4}^{-1}
\nonumber \\
& = 520 \ \left(\frac{1+z}{2}\varepsilon_{peak, 1 \rm{\, MeV}}^{ob.}\right)^2 
\ \Delta t_{-2} \ \Gamma_{2.5}^{-1}   \ \epsilon_{e,-0.5}^{-3}  
\ \epsilon_{B,-0.5}^{-1} \ \theta_{p,0.5}^{-4}. 
\ea
\label{eq:l-eps}
\eeq
Here, $\theta_{p}=10^{0.5}\theta_{p,0.5}$ and $\Delta t=10^{-2}\Delta t_{-2}$~s. Eq.~(\ref{eq:l-eps}) also implies, using Eq.~(\ref{eq:eps_peak}),
\begin{equation}\label{eq:eps-l}
    \varepsilon_{peak}^{ob.} =0.3\frac{2}{1+z} \, {l'}_2^{2/5} \ L_{52}^{1/10} \  \Delta t_{-2}^{-3/5} \ \epsilon_{e,-0.5}^{8/5} \ \epsilon_{B,-0.5}^{1/2} \ \theta_{p,0.5}^{2} \ {\rm MeV},
\end{equation}
where ${l'}=10^2{l'}_2$. Eq.~(\ref{eq:eps-l}) implies that emission peaking at $\sim1$~MeV may be obtained with small compactness, $l'\sim1$, only for very short variability time, $\Delta t\le10^{-4.5}$~s, and, using Eq.~(\ref{eq:l-eps}), large $\Gamma$, $\Gamma\ge10^3$. For the longer variability time commonly assumed in modelling GRBs, $\Delta t\sim1$ to 10~ms \citep[e.g.][]{fireballs1,fireballs2,fireballs3}, $l'\gg1$ is obtained for the parameter range where synchrotron emission peaks at $\sim1$~MeV. The main goal of the present analysis is to examine the modification of the spectrum due to the formation of pairs in moderate to large compactness GRB plasma.

The following point should be noted here. $\Delta t$ in the range of $\sim1$ to 10~ms is commonly adopted, since $\sim1$~ms variability has been observed in some bursts \citep{Bhat92,Fishman94}, and most bursts show variability on $\sim10$~ms time scale \citep{Woods95,Walker00}. It should be kept in mind, however, that variability on much shorter time scale would not have been possible to resolve experimentally, and can not therefore be ruled out. 

Large $l'$ implies large optical depth to photon-photon pair production, $\tau_{\gamma \gamma} \approx l' >1$, and also large optical depth to Thomson scattering by pairs, $\tau_\pm$. In the absence of pair annihilation, $\tau_\pm\approx 2l'$. For $l'\gg1$, $\tau_\pm$ is expected to be large, implying also that pair annihilation is important, since the cross section for pair annihilation is similar to $\sigma_T$ ($v\sigma_\pm(v)\sim c\sigma_T$ for sub-relativistic relative velocity $v$). $\tau_\pm$ may be estimated in this case as follows. As we show in \S~\ref{sec:results}, photons and pairs approach in the case of $l'\gg1$ a quasi-thermal distribution with mildly relativistic effective temperature (or characteristic energy), $\theta m_e c^2$ with $\theta\ll 1$. Under these conditions, the production of pairs via photon annihilation, which for $l'\gg1$ occurs on a time scale much shorter than the dynamical time and may therefore be approximated as the rate of energy production (per unit volume) in $>m_e c^2$ photons, $\sim(\epsilon_e L/4\pi r_i^2\Gamma^2 c)/(m_e c^2t_{dyn})$, is balanced by pair annihilation, the rate of which is given by $\sim n_\pm^2c\sigma_T$. This implies $n_\pm\sim l'^{1/2}/\sigma_T c t_{dyn}$ and 
\begin{equation}\label{eq:tau_pm}
    \tau_\pm\approx l'^{1/2}.
\end{equation}

If synchrotron photons (of energy lower than the pair production threshold $\sim m_e c^2$) and pairs reach a quasi-thermal distribution via Compton and inverse-Compton scattering interactions, then the pair "effective temperature" may be estimated as follows. The energy of low energy photons is increased over a dynamical time by a factor $\simeq\exp(4\tau_\pm\theta)$ (note, that due to plasma expansion the number of scatterings is $\tau_\pm$ rather than $\tau_\pm^2$). The energy $\varepsilon_0$ of the lowest energy photons which reach "thermalization" is therefore given by $\varepsilon_0\exp(4\tau_\pm\theta)\approx\theta m_e c^2$. Assuming that $\varepsilon_0>\varepsilon_{peak}$ and that the synchrotron spectrum is flat, $\varepsilon^2 dN/d\varepsilon\propto\varepsilon^0$, the average energy per photon for synchrotron photons in the energy range $\varepsilon_0<\varepsilon< m_e c^2$ is $\simeq\varepsilon_0\log(m_e c^2/\varepsilon_0)$. Since the number of photons is conserved in Compton scattering interactions, and since the number of pairs is much smaller than the number of photons ($\tau_\pm\approx l'^{1/2}$), conservation of energy implies $\varepsilon_0\log(m_e c^2/\varepsilon_0)\simeq\theta m_e c^2$. Using the relation $\varepsilon_0\exp(4\tau_\pm\theta)\approx\theta m_e c^2$, we therefore find $4\theta\tau_\pm\approx\log[4\theta\tau_\pm-\log(\theta)]$, which implies $4\theta\tau_\pm\sim2$ over a wide range of values of $\tau_\pm\gg1$. The observed effective temperature, $\Gamma\theta$, is therefore
\begin{equation}\label{eq:theta}
    (1+z)^{-1}\Gamma\theta m_e c^2\approx(1+z)^{-1}\frac{\Gamma}{2l'^{1/2}}m_e c^2=5\frac{2}{1+z}\left(\frac{L_{52}\epsilon_{e,-0.5}}{\Delta t_{-4}}\right)^{1/5}l'^{-7/10}_2\,{\rm MeV}.
\end{equation} 
For large optical depth, $\tau_\pm\gg1$, the plasma expands before photons escape. Assuming that the electrons and photons cool "adiabatically", i.e. that the characteristic energy $\theta\propto V^{-1/3}$ where $V$ is the specific volume, the characteristic energy of escaping photons is lower than given by Eq.~(\ref{eq:theta}) by a factor $\tau_\pm^{-1/2}$ (since the optical depth drops as $V^{-2/3}$). The observed characteristic photon energy is therefore
\begin{equation}\label{eq:theta_obs}
    \varepsilon^{ob.}\approx(1+z)^{-1}\frac{\Gamma\theta m_e c^2}{\tau_\pm^{1/2}}\approx(1+z)^{-1}\frac{\Gamma}{2l'^{3/4}}m_e c^2=2\frac{2}{1+z}\left(\frac{L_{52}\epsilon_{e,-0.5}}{\Delta t_{-4}}\right)^{1/5}l'^{-19/20}_2\,{\rm MeV}.
\end{equation} 

In the limit of $l'\rightarrow\infty$ we expect the plasma to reach thermal equilibrium. Assuming that the fraction of dissipated kinetic energy carried by electrons is converted to thermal radiation, the resulting (blue shifted) radiation temperature is
\begin{equation}\label{eq:T}
\Gamma T = 0.1(1+z)^{-1} \, {l'}_2^{1/10} \ L_{52}^{3/20} \ \Delta t_{-4}^{-2/5} \ 
\epsilon_{e,-0.5}^{-1/10} \ {\rm MeV}.
\end{equation}

\section{Numerical calculations}
\label{sec:numerical}

\subsection{Method}

The acceleration of particles in the internal shock waves is accompanied by time dependent radiative processes, which are coupled to each other. In order to follow the emergent spectra we developed a time dependent model, solving the kinetic equations describing cyclo-synchrotron emission, synchrotron self absorption, Compton scattering ($e \gamma \rightarrow e\gamma$), pair production ($e^+e^-\rightarrow \gamma \gamma$) and annihilation ($ \gamma \gamma \rightarrow e^+e^-$). Our model follows the above mentioned phenomena over a wide range of energy scales, including the evolution of the rapid electro-magnetic cascade at high energies.

The calculations are carried out in the comoving frame, assuming homogeneous and isotropic distributions of both particles and photons in this frame. Relativistic electrons are continuously injected into the plasma, at a constant rate and with a pre-determined constant power law index $p$ between $\gamma_{min}$ and $\gamma_{max}$ (see eq. \ref{eq:gamma_min}), throughout the dynamical time, during which the shock waves cross the colliding shells.  Above $\gamma_{max}$, an exponential cutoff is assumed. The magnetic field is assumed to be time independent, given by equation \ref{eq:B}. 

The particle distributions are discretized, spanning a total of 10 decades of energy, ($\gamma \beta_{min} = 10^{-3}$ to $\gamma \beta_{max} = 10^{7}$). The photon bins span 14 decades of energy, from $x_{min} \equiv \varepsilon_{min}/m_ec^2 = 10^{-8}$ to $x_{max} \equiv \varepsilon_{max}/m_ec^2 = 10^{6}$. A fixed time step is chosen, typically $10^{-4.5}$ times the dynamical time. Numerical integration, using Cranck-Nickolson second order scheme for synchrotron self absorption and first order integration scheme for the other processes, is carried out with this fixed time step. At each time step we calculate (i) The energy loss time of electrons and positrons at various energies (via cyclo-synchrotron emission and inverse Compton scattering taking into account the fact that low energy electrons can gain energy via direct Compton scattering); (ii) The annihilation time of electrons and positrons; (iii) The annihilation and energy loss time of photons. Electrons, positrons and photons for which the energy loss time or annihilation time are smaller than the fixed time step, are assumed to lose all their energy in a single time step, producing secondaries which are treated as a source of lower energy particles. The calculation is repeated with a shorter time steps, until convergence is reached.

In the calculations, the exact cross sections for each physical phenomenon, 
valid at all energy range are being used.
Calculations of the reaction rate and emergent photon spectrum from Compton scattering 
follow the exact treatment of \citet{Jones68}. 
Pair production rate, and spectrum of the emergant pairs are calculated using the results of 
\citet{BS97}. Pair annihilation calculations are carried out using the exact cross section first
derived by \citet{Svensson82b}.
The power emitted by an electron with an arbitrary Lorentz factor $\gamma$ in a magnetic field, is 
calculated using the cyclo-synchrotron emission pattern 
\citep[see][]{Bekefi66, Ginzburg69, Mahadevan96}.

A full description of the physical processes, 
the kinetic equations solved and the numerical methods used appear in
\citet{Pe'er03b}.

\subsection{Adiabatic expansion}
\label{sec:adiabatic}

Once the two shocks cross the colliding shells, the dissipation of
kinetic energy ceases, and the compressed shells expand and cool. The
thermal energy carried by protons, electrons (positrons) and magnetic
field decreases and converted back to kinetic energy. If the
scattering optical depth at the end of the dynamical time is small,
photons escape the shells and the energy they carry is "lost" from the
plasma. If the optical depth is large, then photons interact with the
expanding electrons and positrons, and this interaction affects the
emerging spectrum. Since the plasma is collisionless, and particles
are coupled through macroscopic electro-magnetic waves, the details of
the conversion of thermal to kinetic energy are unknown. For
relativistic plasma at thermal equilibrium undergoing adiabatic
expansion, the pressure is inversely proportional to   
$V^{4/3}$, where $V$ is the volume. We assume that the plasma expands
in the comoving frame with velocity comparable to the adiabatic sound
speed, $c/\sqrt{3}$, that $B \propto V^{-2/3} \propto t^{-2}$ and that
electrons and positrons lose energy due to expansion at a rate
$d\varepsilon/\varepsilon=-dV/3V$.  

Since our numerical model is spatially "0-dimensional", we calculate
the evolution of the photon and particle spectrum in the scenario
outlined in the previous paragraph, assuming a uniform isotropic
particle and photon distribution. This calculation does not take into
account the energy loss of photons due to the bulk expansion velocity
of the electrons, which implies that photons are more likely to
collide with electrons that move away from (rather than towards)
them. In order to estimate the effect of this energy loss, we have
carried out the following calculations. 

\begin{figure}
\plotone{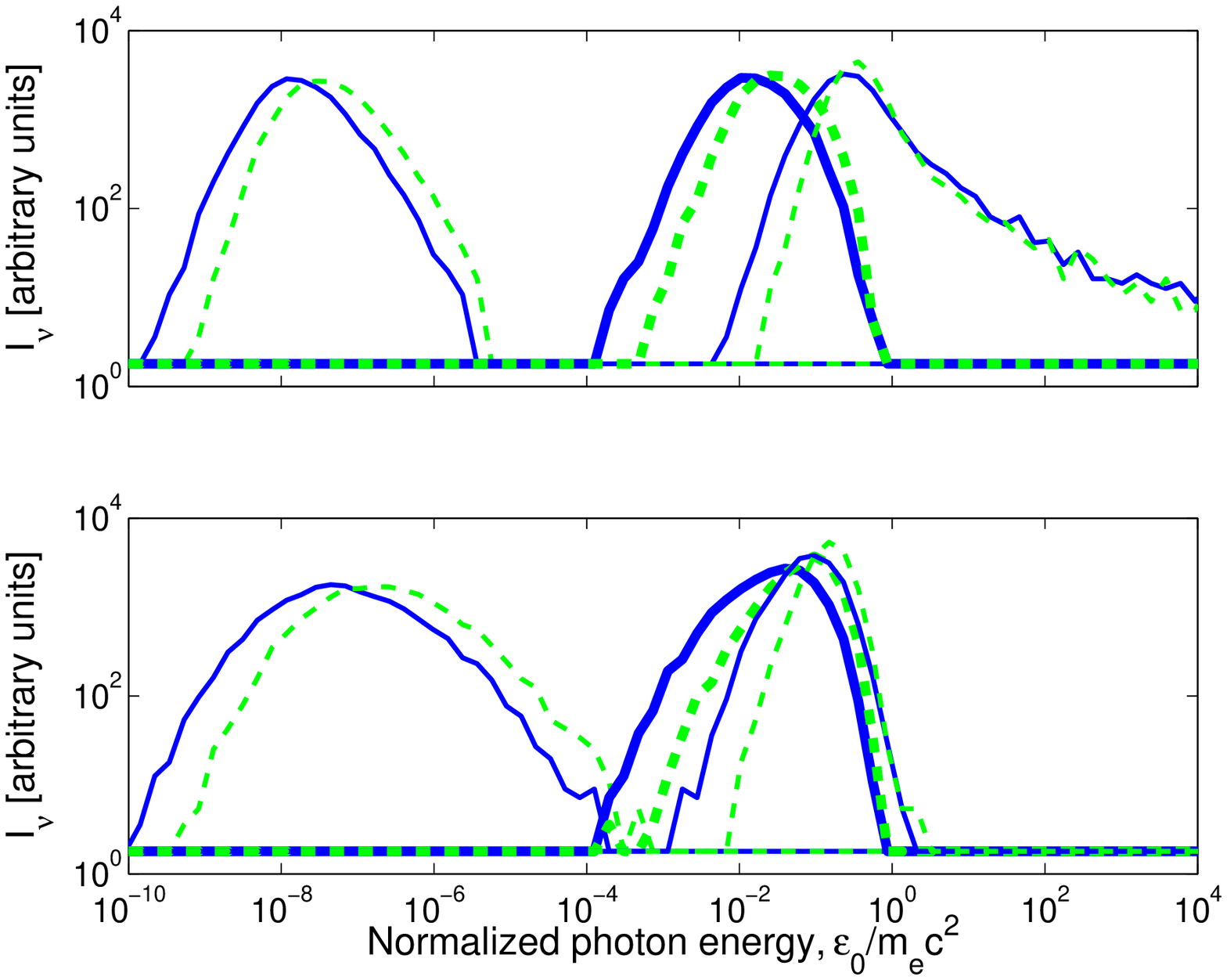}
\caption{Monte-Carlo calculations of photon spectra emerging following the injection of  
mono-energetic photons at the center of an adiabatically expanding sphere of electrons with initial temperature $\theta \equiv kT/m_ec^2 = 0.1$ (see \S~\ref{sec:adiabatic}).
Upper panel: initial scattering optical depth $\tau = 10$, lower panel: $\tau=25$.
Results are shown for several initial photon energy (normalized to $m_ec^2)$:
$\varepsilon_0 = 10^{-8}$ (left, thin), 
$\varepsilon_0 = 10^{-2}$ (intermediate, bold),
$\varepsilon_0 = 10^{4}$ (right, thin).
Solid lines show the results of complete calculations, while 
dashed lines show results of calculations where energy loss of photons due to the bulk motion of electrons are neglected.
}
\label{fig:MonteCarlo}
\end{figure}

\begin{figure}
\plotone{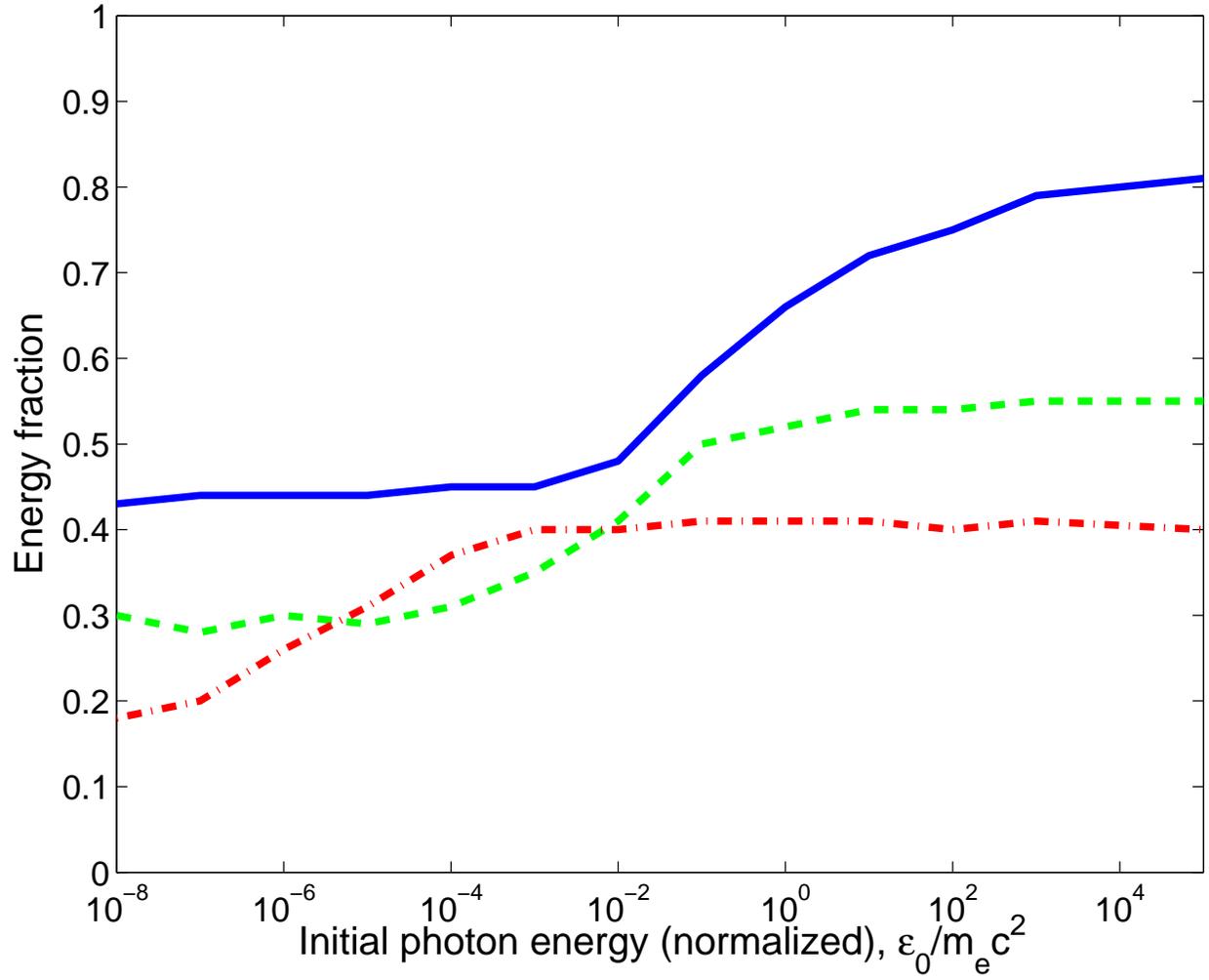}
\caption{The ratio of the average energy of photons emerging from an expanding sphere  (see Fig.~\ref{fig:MonteCarlo} caption) obtained
when energy loss to bulk motion is included and excluded.
Results shown for several
initial optical depths: $\tau = 10$ (solid),
$\tau=25$ (dashed), $\tau=100$ (dash-dotted).
}
\label{fig:ad_expansion}
\end{figure}

We have calculated, using Monte-Carlo technique, the evolution of the
momentum of a mono-energetic beam of photons emitted at the center of
an expanding spherical ball of thermal electron plasma, until they
escape. The plasma ball was assumed to expand with radial velocity
$v(r) =[r/R(t)]c/\sqrt{3}$, where $R(t)$ is the ball radius, and its
temperature was assumed to decrease as $\theta \propto R^{-1}$ from an
initial value of $\theta=0.1$. The emergent photon spectrum is shown
in figure~\ref{fig:MonteCarlo} for three different initial photon
energies,  
$\varepsilon_0/m_ec^2 = 10^{-8}, 10^{-2}, 10^4$, and two initial
scattering optical depths, $\tau = 10, 25$. These calculations were
repeated omitting the energy loss of the photons due to the bulk
motion of the electrons, by assuming $v=0$ (while keeping the ball
expansion and temperature decrease unchanged). Figure
\ref{fig:ad_expansion} shows the ratio of the average energy of
emerging photos with and without inclusion of energy loss to bulk
motion for several initial optical depths, as a function of the
initial photon energy. This figure demonstrate that the effect of
energy loss to bulk expansion is not highly dependent on the initial
photon energy, and that it leads to reduction of photon energy by a
factor $\sim3$ for initial optical depths in the range of 10 to 100. 

In the numerical calculations presented in \S~\ref{sec:results} and
\S~\ref{sec:ghis} we have corrected for the effect of energy loss due
to bulk motion by multiplying the emerging photon energy by the
(energy dependent) factor inferred from the calculations presented in
figure~\ref{fig:ad_expansion}. This correction is applied for photons
of energy exceeding the self absorption energy $\varepsilon_{ssa}$ at
the beginning of the expansion phase. For photons of energy lower than
the self absorption energy $\tilde{\varepsilon}_{ssa}$ at the end of
the expansion phase, where the optical depth drops to unity, we have
applied no correction. This is due to the fact that photons at these
energies are tightly coupled to the electrons, they are continuously
emitted and absorbed, and this coupling is the dominant factor
determining the photons' energy. Note, that the self absorption energy
decreases during the adiabatic expansion, as the density and the
magnetic field decrease. 
At the intermediate energy range, $\varepsilon_{ssa}$ to
$\tilde{\varepsilon}_{ssa}$, we have applied an interpolated
correction factor.  

Figure \ref{fig:adiabatic1} 
presents an example of the modification of the spectrum due to bulk
expansion. Since the fractional energy loss is not strongly energy
dependent, the correction we apply does not lead to significant
distortion of the spectrum. Note, however, that since we do not take
into account the spread in the energy of emerging photons that
initially had the same energy, see figure~\ref{fig:MonteCarlo}, but
rather apply a single correction factor appropriate for the average
energy of emerging photons, the emerging spectrum would in reality be
somewhat "smeared" compared to our calculation. In particular, the
annihilation peak that appears at $\sim 100 \MeV$ is expected to be
"smoothed".

\section{Results}
\label{sec:results}

We have shown in \S~\ref{sec:compactness} that $l'\gg1$ is obtained for the parameter range where synchrotron emission in the fireball model peaks at $\sim1$~MeV [see  Eq.~(\ref{eq:eps-l})]. In this section we present detailed numerical calculations investigating the emission of radiation at moderate to large compactness (for completeness, we present in \S~\ref{sec:low} numerical results also for low compactness). We have demonstrated in \S~\ref{sec:compactness} that for large compactness, the characteristics of emitted radiation are determined mainly by $l'$, with weak dependence on the values of other parameters [see, e.g., Eq.~(\ref{eq:theta_obs})]. The results presented in \S~\ref{sec:intermediate} and \S~\ref{sec:high} for particular choices of parameter values (e.g. $\Delta t=10^{-3}$~s and $\Delta t=10^{-4}$~s) with $l'\sim10^2$ to $10^3$, are therefore expected to characterize the emission also for other choices of parameters with similar values of $l'$. 

\subsection{Low compactness}
\label{sec:low}

Figure \ref{fig:results1} shows spectra obtained for low compactness,
$l' \lesssim 3$. 
Synchrotron self absorption results in quasi-thermal spectrum at low 
energies, below $\varepsilon_{ssa} \approx 100 \eV- 1\keV$. Between
$\varepsilon_{ssa}$ and $\varepsilon_{peak} \approx 10 - 100 \keV$,
the spectral index is softer than expected for synchrotron emission
only: 
$\nu F_\nu\propto\nu^\alpha$ with $\alpha \approx 0.3$
rather than $\alpha \approx 0.5$.
The reason for this deviation is related with the particle
population, presented in Figure~\ref{fig:elec}. 
The self absorption phenomenon causes particles to accumulate at low to
intermediate energies, forming a quasi-Maxwellian distribution with
temperature $\theta \approx 0.5$. Therefore, above this energy, the
electron spectral index is somewhat softer than the spectral index
expected without  the self absorption phenomenon, 
($d n_e/ d\varepsilon_e \propto \varepsilon_e^{-2.4}$  instead of 
$d n_e/ d \varepsilon_e \propto \varepsilon_e^{-2}$),
resulting in a steeper slope above $\varepsilon_{ssa}$.

Between  $\varepsilon_{peak} \approx 10- 100 \keV$ and $10 \MeV$
the spectral slope $\alpha \approx -0.3$  is harder than expected
($\alpha = -0.5$ for $p=3.0$) due to significant inverse Compton emission.
The combined effects, of relatively soft spectrum at low energies, and
cooling of particles by both synchrotron emission and Compton
scattering lead to the creation of a very soft spectrum
($\alpha \approx 0.1$) near the IC
peak at high  energies ($3- 30 \GeV$). 
It is therefore concluded, that for low compactness, $l' \lesssim 3$,
and $\epsilon_B \simeq \epsilon_e$,
the spectrum expected at all energy bands, between $\sim 100 \eV - 30
\GeV$ is flat, with spectral index that varies in the range
 $-0.3 \lesssim \alpha \lesssim +0.3$.

The flattening of the spectrum due to both synchrotron and IC scattering
decreases the dependence on p. 
As presented in Figure~\ref{fig:p1},
for $p=2.0$ the flux rises slowly at all energy bands due to IC scattering,
while for $p=2.5$ it is nearly constant in the  
1 \keV- 1 \GeV range. 

The dependence of the spectrum on magnetic field equipartition fraction is
shown in Figure~\ref{fig:xi_B}, which demonstrates that 
comparison of the flux at 1~\GeV and 100~\keV 
may allow to determine the value of $\epsilon_B$.

\subsection{High compactness}
\label{sec:high}

\begin{figure}
\plotone{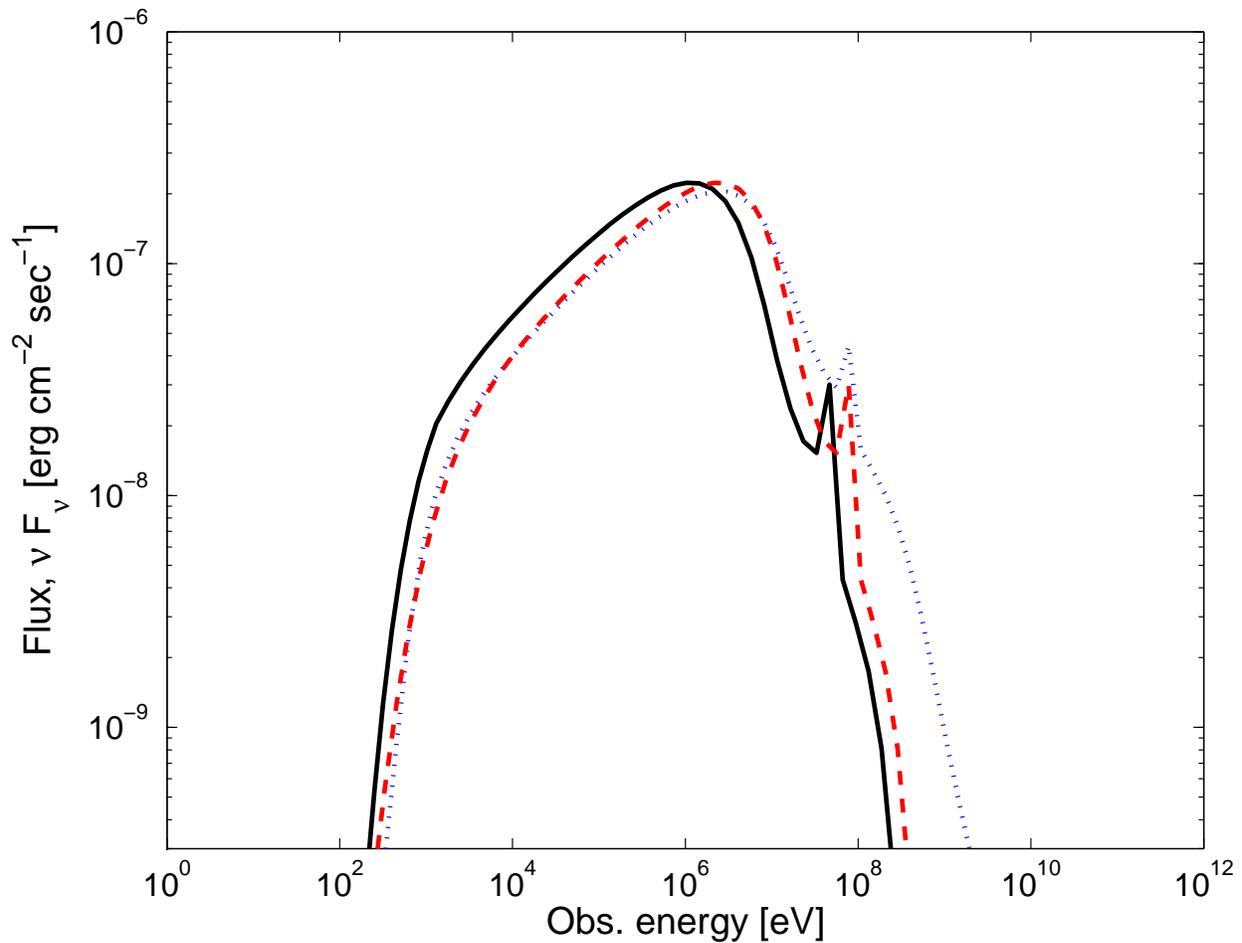}
\caption{The effect of adiabatic expansion on emergent spectra in the case of large compactness ($l'=250$): Dotted: spectrum at the end of the dynamical time 
(before adiabatic expansion); Dashed: spectrum after adiabatic expansion;
Solid: after correction for energy loss due to plasma bulk motion (see \S~\ref{sec:adiabatic}).
}
\label{fig:adiabatic1}
\end{figure}

Figure \ref{fig:results2} presents results for large compactness.
When the compactness is large enough, Compton scattering by pairs 
becomes the dominant emission mechanism. 
The spectrum can not be approximated in this case by the 
commonly used optically thin synchrotron-self-Compton model.
As demonstrated in Figure~\ref{fig:elec},
electrons and positrons lose their energy much faster than
the dynamical time scale, and quasi Maxwellian distribution
with an effective temperature $\theta \simeq 0.05 - 0.1$ is formed.
The energy gain of the low energy 
electrons by direct Compton scattering 
results in a spectrum steeper than Maxwellian at the low energy end,
indicating that a steady state did not develop.
As shown in Figure~\ref{fig:elec_ad}, the electron distribution approaches 
a Maxwellian at the end of the adiabatic expansion. 
The self absorption frequency  
$\varepsilon_{ssa}^{ob.} \approx 3 - 10 \keV$ {\it before} the adiabatic
expansion (see Figure \ref{fig:adiabatic1}) 
is well approximated by equation (\ref{eq:eps_ssa_thermal}), valid
for thermal distribution of electrons.

The ratio of pair to proton number density at the end of the dynamical time is 
$f \equiv n_{\pm}/n_p \simeq 10$ in the calculations shown in
Figure~\ref{fig:results2}, 
in agreement with the analytical approximations of \S~\ref{sec:compactness}.
This ratio is determined by balance between pair production and pair annihilation,
and leads to optical depth $\tau_\pm \simeq l'^{1/2}$ (Figure \ref{fig:tau}), in accordance with the predictions of \S~\ref{sec:compactness}.
Pair annihilation phenomenon creates the peak at $\Gamma m_e c^2 \sim 10^2 \ \Gamma_{2.5} \MeV$ for large compactness.

Intermediate number of scattering $\tau_\pm \lesssim 20$,
as is the case for $\Delta t = 10^{-4} \rm{\ s}$ 
(Figures \ref{fig:adiabatic1}, \ref{fig:tau}), leads to 
moderate value of the Compton $y$ parameter.
If the electrons distribution is approximated as a Maxwellian
with temperature $\theta \approx 0.1$ (see Figure \ref{fig:elec}), 
the Compton $y$ parameter,
$y \simeq 4 \theta \tau \approx 4 \ \theta_{-1} \tau_{1}$
is not much higher than 1.
In this scenario, the number of scatterings is not large enough to create 
a $\nu F_\nu \propto \nu^1$ spectrum, 
expected when Comptonization is the dominant emission mechanism
(note, that the observed spectral indices at $1-10 \keV$ 
reported by \citet{Crider,Frontera} and \citet{Preece} 
are even harder than this value).  
The resulting spectrum has a slope  
 $\nu F_\nu \propto \nu^\alpha$ with $\alpha \approx 0.5$
between $\varepsilon_{ssa} \approx 3-10 \keV$
and $\varepsilon_{peak} \approx 10 \MeV$.
For $\Delta t = 10^{-5} \rm{\ s}$,  
the optical depth is $\tau_\pm \approx 60$ and the 
Compton y-parameter is higher, $y \approx 25$,
resulting in a steeper slope.
However, even in this scenario the slope is $\alpha \approx 0.7$ and not 
the limiting value, $\alpha = 1$.

Since the dominant emission mechanism is Compton scattering
by the quasi-thermally distributed particles,
the spectrum is independent on
most of the physical parameters related with particle acceleration. 
As is seen in Figure \ref{fig:p3}, the spectra does not depend 
on the power law index of the accelerated
electrons.
The dependence on $\epsilon_B$ is weak; smaller $\epsilon_B$ gives a
steeper slope between $1 \keV - 10 \MeV$.

If $l'$ is large, the flux is rising up to $1 \MeV$, 
regardless of the value of $p$. Therefore, observing a decrease in the flux between $30 \keV$ - $1 \MeV$ indicates both $p>2$ 
and $l' \lesssim 3$.

\subsection{Intermediate compactness}
\label{sec:intermediate}

For $l'$ in the range of few to few tens, synchrotron
emission and Compton scattering equally contribute to the observed
radiation. Here too, the spectrum cannot be approximated by simple
analytic formulation. 
Figure~\ref{fig:results3} shows several examples of spectra obtained
for moderate $l'$.  

Even though a significant number of pairs are created, $f \sim10$ for
the three scenarios presented in figure~\ref{fig:results3}, the
optical depth for scattering is approximately 1, and the self
absorption frequency is $\varepsilon_{ssa} \sim 1 \keV$. 
Although the number of scatterings is not large, it is sufficient for
increasing the energy at which the spectrum peaks by about a factor of
3 above the synchrotron  
model prediction, leading to $\varepsilon_{peak} \sim 500 \keV$ for
$\epsilon_e$ and $\epsilon_B$ near equipartition. 
Lower values of $\epsilon_B$ lead to higher $\varepsilon_{peak}$,
while lower values of $\epsilon_e$ lead to lower
$\varepsilon_{peak}$. 

The spectral slope in the $1 \keV - 1 \MeV$ range 
is soft $\alpha \simeq 0.3$, instead of the expected value $\alpha \approx
0.5$, similar to the spectra produced for small $l'$,
due to similar reasons.
Figure \ref{fig:p2} shows that the exact power law 
of the accelerated electrons has only a minor influence on
the observed spectra.

A significant flux is expected up to $\gtrsim 1 \GeV$.
Unlike the scenario of very small $l'$, the flux decreases
above $\varepsilon_{peak}$, and no second peak due to IC scattering
from high energy electrons is formed. 
This phenomenon is due to pair production,
that cuts the flux above the energies where such a peak
would form, at $\sim 1 \GeV$.
Therefore, the main characteristics of spectra produced by
intermediate values of $l'$ 
are a moderate increase in the flux in the 1~\keV to 1~\MeV energy bands,
a moderate decrease in the flux in the 
1 \MeV to 1 \GeV band, and a sharp cutoff above this energy.

\section{Quasi thermal Comptonization}
\label{sec:ghis}

We consider in this section the quasi-thermal Comptonization scenario proposed by  \citet{Ghisellini99}. This model is different than the one considered in the previous section in that (i) It is assumed that the kinetic energy dissipated in two-shell collision is continuously distributed among all shell electrons, rather than being deposited at any given time into a small fraction of the shell electrons that pass through the shock wave, and (ii) It is assumed that energy is equally distributed among electrons, rather than following a power-law distribution. 
In this case, no highly relativistic electrons are produced, and Comptonization is the main mechanism responsible for emission above a few keV:
Synchrotron emission is self absorbed up to frequency $\nu_T$,
providing the seed photons for Comptonization 
creating a $\nu F_\nu \propto \nu^1$ spectrum up to $\varepsilon_{peak}$.
We derive here the constraints on the peak flux energy $\varepsilon_{peak}$ 
in this scenario, both analytically and numerically.

Figure~\ref{fig:ghisellini} presents numerical results obtained for this scenario. The calculations were carried out using a modified version of the numerical model,
in which electrons and positrons are forced to follow a Maxwellian distribution, 
$n(\gamma) d\gamma \propto \beta \gamma^2 e^{-\gamma/\theta} d\gamma$. 
The temperature $\theta(t)$ is determined self-consistently by the balance of energy injection and energy loss. The results are shown for two representative cases in Figure~\ref{fig:ghisellini}.
In both cases, electrons and positrons reach a mildly relativistic temperature, $\theta\sim0.3$, and the spectrum peaks at $\varepsilon_{peak} \approx 30 \MeV$. 
The spectral slope below $\varepsilon_{peak}$ is $\alpha \approx 0.5$ ($\nu F_\nu \propto \nu^\alpha$) for $\Delta t = 10^{-3}\rm{\ s}$, where $\tau_\pm \sim 2$, and $\alpha \approx 1$ for $\Delta t = 10^{-4}\rm{\ s}$, where $\tau_\pm \sim 10$. In the former case, the scattering optical depth, $\tau_\pm \sim 2$, is not large enough to produce the $\alpha \simeq 1$ spectrum expected for $\tau_\pm\gg1$. We show below, using analytic analysis, that $\varepsilon_{peak} \gtrsim 30 \MeV$ is a generic result of the thermal-Comptonization scenario.

In an expanding plasma characterized by width $\Delta R$, 
the available time for scattering is $\Delta R / c$, 
and the time between scattering is $\langle l \rangle/c$, 
where $\langle l \rangle$ is the mean free path.  
Therefore, the number of scattering,
$N \approx \Delta R / \langle l \rangle$, is proportional to $\tau$,  
and the Compton $y$ parameter is given by
\beq
y = 4 \theta \tau = 4 \theta \Delta R \sigma_T n_p f = 
4\theta \Gamma c \Delta t \sigma_T n_p f.
\label{eq:y}
\eeq
The synchrotron spectrum is well approximated by a black body
radiation up to frequency $\nu_T$, given in equation \ref{eq:nu_T}.
Thus, the observed luminosity is given by
\beq
\frac{L}{\Gamma^2} = e^y L_{sync} =
 e^y \frac{8 \pi}{3} m_e r_i^2 \theta \nu_T^3,
\label{eq:L}
\eeq
and the observed emission peaks at
\beq
\varepsilon_{peak}^{ob.} = e^y h \nu_T \Gamma. 
\label{eq:eps_peak_ghis}
\eeq
In the following calculation, we assume that $e^y h \nu_T \Gamma$
is not larger than the saturation value of $\varepsilon_{peak}^{ob.}$, 
$4 \theta \Gamma m_e c^2$.
If this is not the case, for $\Gamma = 10^{2.5}$, $\theta \approx 10^{-3}$ is needed 
in order to obtain $\varepsilon_{peak}^{ob.} \approx 1 \MeV$.
Since $y > few$ is required, this value of $\theta$ implies a very large 
optical depth, $\tau \approx 10^3$. Such large value is not obtained for parameter values relevant for GRB fireballs, and would lead to strong suppression of emitted radiation.

Assuming that $\varepsilon_{peak}^{ob.}$ is given by equation~\ref{eq:eps_peak_ghis},
since $\theta \leq 1$, in order to get $y > few$ 
photons have to undergo a minimum
number of scatterings, i.e. $\tau \gtrsim few$.
For a minimum value of $\tau = 10^{0.5} \tau_{0.5}$, 
using $\tau = \Delta R \sigma_T n_p f$, $f$ is given by
\beq
f = 4 \quad L_{52}^{-1} \ \Gamma_{2.5}^5 \ \Delta t_{-4} \tau_{0.5}.
\label{eq:f}
\eeq
Eliminating $e^y$ from equations \ref{eq:L}, \ref{eq:eps_peak_ghis} 
using equation \ref{eq:nu_T}, $\theta$ is given by
\beq
\theta = 1.7 f^{-1/9} \quad
 L_{52}^{-2/30} \epsilon_{B,-0.5}^{-0.3} 
{\varepsilon_{peak,1\rm{\,MeV}}^{ob.}}^{-1/3} \Gamma_{2.5}^{5/9}, 
\label{eq:theta1}
\eeq
and the Compton $y$ parameter (eq. \ref{eq:y}) becomes
\beq
y = 4 \theta \tau = 5.5 f^{8/9} \quad
 L_{52}^{14/15} \epsilon_{B,-0.5}^{-0.3} \Gamma_{2.5}^{-40/9} 
\Delta t_{-4}^{-1}  {\varepsilon_{peak,1\rm{\,MeV}}^{ob.}}^{-1/3}.
\label{eq:y2}
\eeq
Using the values obtained for $\nu_T$ (eq. \ref{eq:nu_T}) 
and $\theta$ (eq. \ref{eq:theta1}) in 
equation \ref{eq:eps_peak_ghis}, we obtain
\beq
e^y = \frac {\varepsilon_{peak}}{ h \nu_T \Gamma} = 135 f^{-1/18} 
\quad L_{52}^{-8/15} \epsilon_{B,-0.5}^{-0.15} 
{\varepsilon_{peak,1\rm{\,MeV}}^{ob.}}^{4/3} 
\Gamma_{2.5}^{16/9} \Delta t_{-4}.
\eeq
The very weak dependence of the right hand side on $f$, allows eliminating it.
Taking $f^{1/18} \approx 1$ one can take the log of both sides and 
use equation \ref{eq:y2} to express $f$, 
\beq
f \approx 0.55 \quad L_{52}^{-21/20} \epsilon_{B,-0.5}^{27/80} 
\Gamma_{2.5}^5 \Delta t_{-4}^{9/8} 
{\varepsilon_{peak,1\rm{\,MeV}}^{ob.}}^{3/8}.
\label{eq:f_final1}
\eeq
Equation \ref{eq:f} then gives
\beq
\varepsilon_{peak}^{ob.} = 
200 (1+z)^{-1} \quad L_{52}^{2/15} \epsilon_{B,-0.5}^{-9/10} 
\Delta t_{-4}^{-1/3} \tau_{0.5}^{8/3} \, \MeV.
\label{eq:eps_final}
\eeq
Since $\epsilon_B$ already assumes its maximum value, 
the only way to meet the observed peak flux 
$\varepsilon_{peak}^{ob.} \approx 1 \MeV$ 
in this scenario is by assuming a significantly lower value of total luminosity
or longer variability time  $\Delta t$,
both inconsistent with the parameter space region for high compactness,
as well as with the observations. Note, that the observed peak would be obtained at an energy somewhat lower than given by Eq.~\ref{eq:eps_final}, due to pair production which leads to a cutoff at $\Gamma m_e c^2 = 150  \ \Gamma_{2.5} \MeV$.

The above analysis shows that in the "slow heating scenario"
the flux cannot peak below few hundred \MeV. 
This general result is not changed by the adiabatic expansion,
during which the optical depth decreases according to 
$\tau(t)= \tau_0 [\Delta R /(\Delta R + vt)]^2$, 
where $\tau_0$ is the optical depth before the expansion
and $v \approx c/\sqrt{3}$ is 
the expansion velocity. The total number of scattering increases 
during the adiabatic expansion by a factor of $(1+ \sqrt{3}) \approx 3$
compared to the number of scattering at the dynamical time.
This factor enters
equations \ref{eq:f} and \ref{eq:f_final1}
with nearly the same power (1 in equation \ref{eq:f} , 9/8 on equation \ref{eq:f_final1}), leaving the final conclusion unchanged.

\section{Summary \& discussion}
\label{sec:summary}

Within the fireball model framework, synchrotron emission peaking at $\sim1$~MeV may be obtained with small compactness, $l'\sim1$, only for very short variability time, $\Delta t\le10^{-4.5}$~s, and large $\Gamma$, $\Gamma\ge10^3$ [see Eqs.~(\ref{eq:eps-l}) and ~(\ref{eq:l-eps}), of \S~\ref{sec:compactness}]. For the longer variability time commonly assumed in modelling GRBs, $\Delta t\sim1$ to 10~ms \citep[e.g.][]{fireballs1,fireballs2,fireballs3}, $l'\sim10^2$ to $l'\sim10^3$ is obtained for the parameter range where synchrotron emission peaks at $\sim1$~MeV (For smaller compactness, spectra peak at lower, X-ray, energy). This result has two main consequences. First, observed GRB spectra are expected to be significantly modified by the presence of pairs. Second, peak energy $\gg1$~MeV can not be obtained for $L\sim10^{52}{\rm erg/s}$, since it would imply $l'>>10^3$ [see Eqs.~(\ref{eq:eps-l}),~(\ref{eq:T}),~(\ref{eq:theta_obs})], in which case most of the radiation will not escape due to large optical depth to Thomson scattering by pairs. These conclusions are consistent with the conclusions of \citet{GSW01}, and may provide an explanation for the lack of bursts with peak energy $\gg1$~MeV. 

It should be noted at this point that GRB observations do not allow, in most cases, to identify variability on $\sim1$~ms time scale. Rapid variability, on $\sim1$~ms time scale, has been observed in some bursts \citep{Bhat92,Fishman94}, and most bursts show variability on the shortest resolved time scale, $\sim10$~ms \citep{Woods95,Walker00}. It should be kept in mind, however, that variability on much shorter time scale would not have been possible to resolve experimentally, and can not therefore be ruled out. 

We have demonstrated in \S~\ref{sec:compactness} that for $l'$ in the range of $10^2$ to $10^3$ the characteristics of emitted radiation are determined mainly by $l'$, with weak dependence on the values of other parameters [see, e.g., Eq.~(\ref{eq:theta_obs})]. The peak of the specific luminosity is expected to be close to $\sim1$~MeV, $\sim 1 (L_{52}/\Delta_{t,-3})^{1/5}(l'/100)^{-1}$~MeV [Eq.~(\ref{eq:theta_obs})], and the spectrum is expected to differ significantly from an optically thin synchrotron spectrum. 

\begin{figure}
%\figurename{namename}
%\epsscale{1.0}
\plotone{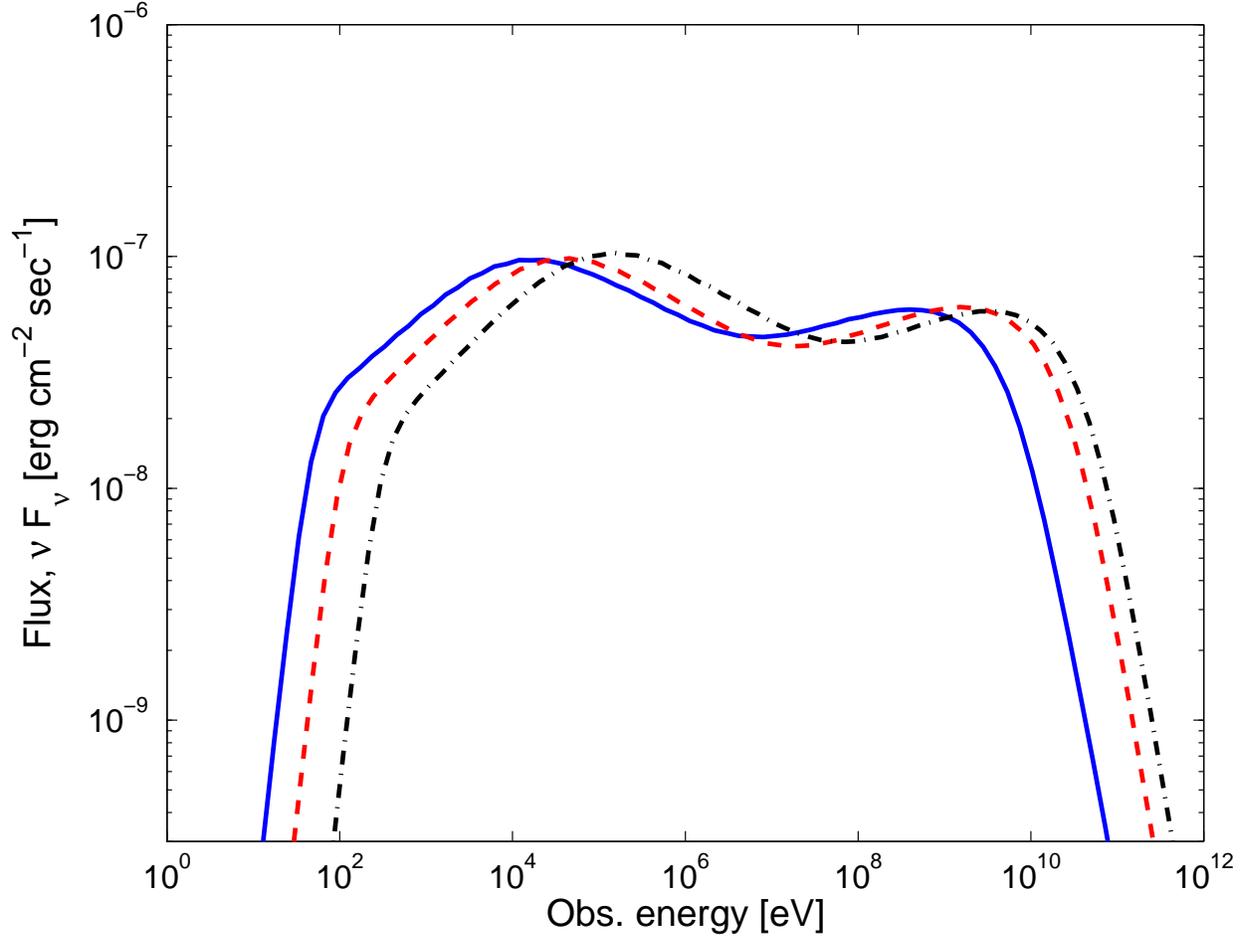}
\caption{Time averaged spectra obtained for low compactness parameter.
Results are shown for $L=10^{52}$~erg, $\epsilon_e=\epsilon_B=10^{-0.5}$, $p=3$ (all cases) and: $\Delta t = 10^{-2} {\rm \ s}$ , $\Gamma = 300$ (solid);
$\Delta t = 10^{-3} {\rm \ s}$ , $\Gamma = 600$ (dashed);
$\Delta t = 10^{-4} {\rm \ s}$ , $\Gamma = 1000$ (dash-dotted).
The compactness parameter is $l' = 2.5, 0.8, 0.6$, respectively.
Luminosity distance $d_L=2 \times 10^{28}$ and $z=1$ were assumed.
}
\label{fig:results1}
\end{figure}

\begin{figure}
\plotone{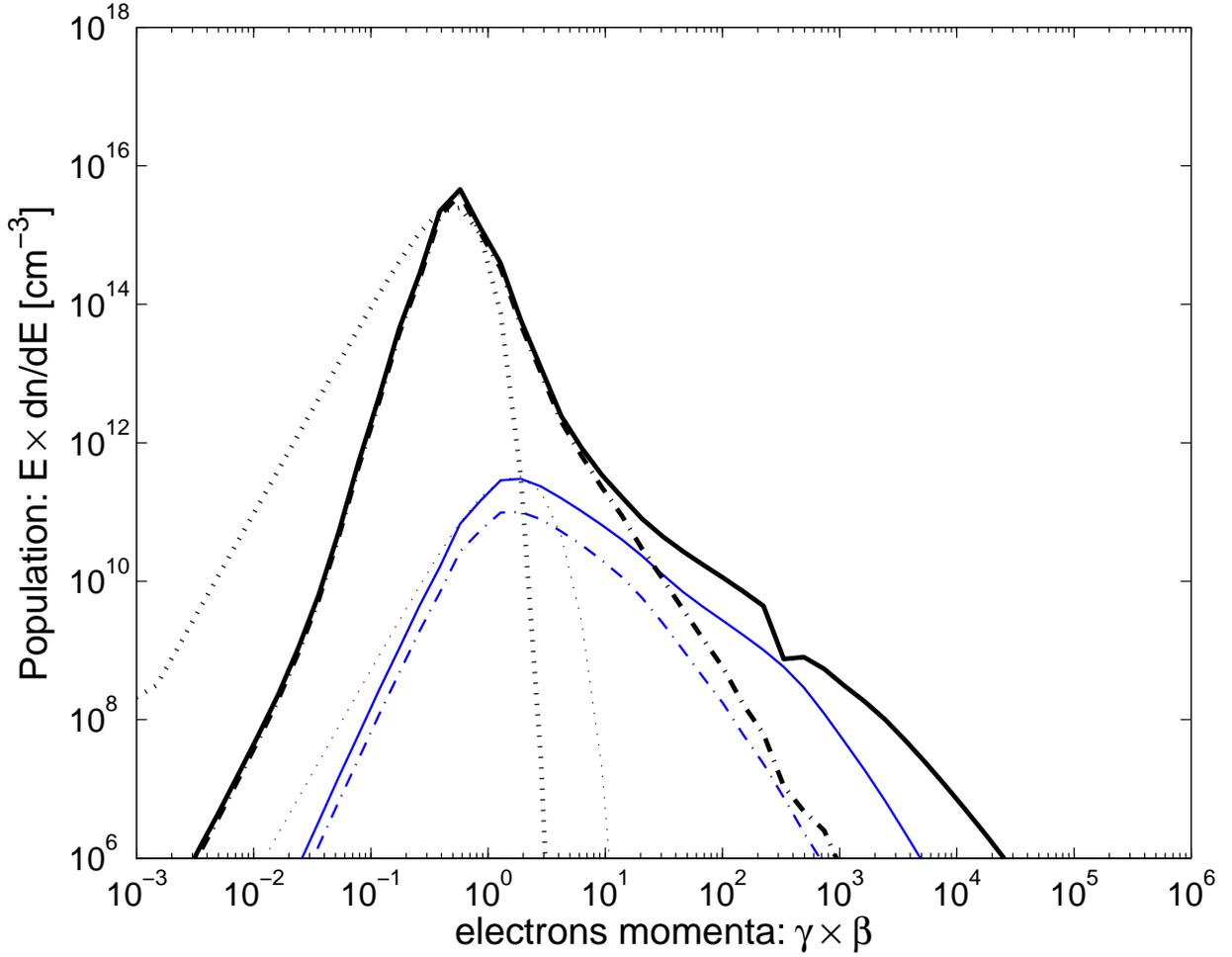}
\caption{Particle distribution at the end of the dynamical time.
Thick:  $\Delta t = 10^{-4} {\rm \ s}, p=3.0$, $\Gamma = 300$, $l'=250$;
Thin: $\Delta t = 10^{-4} {\rm \ s}, p=3.0$, $\Gamma = 1000$, $l'=0.6$.
All other parameters are the same as in Figure \ref{fig:results1}.
Solid: electron distribution, dash-dotted: positron distribution.
The dotted lines show Maxwellian distributions at temperatures 
$\theta\equiv kT/m_ec^2=0.08$ (thick) and
$\theta= 0.5$ (thin).
}
\label{fig:elec}
\end{figure}

\begin{figure}
\plotone{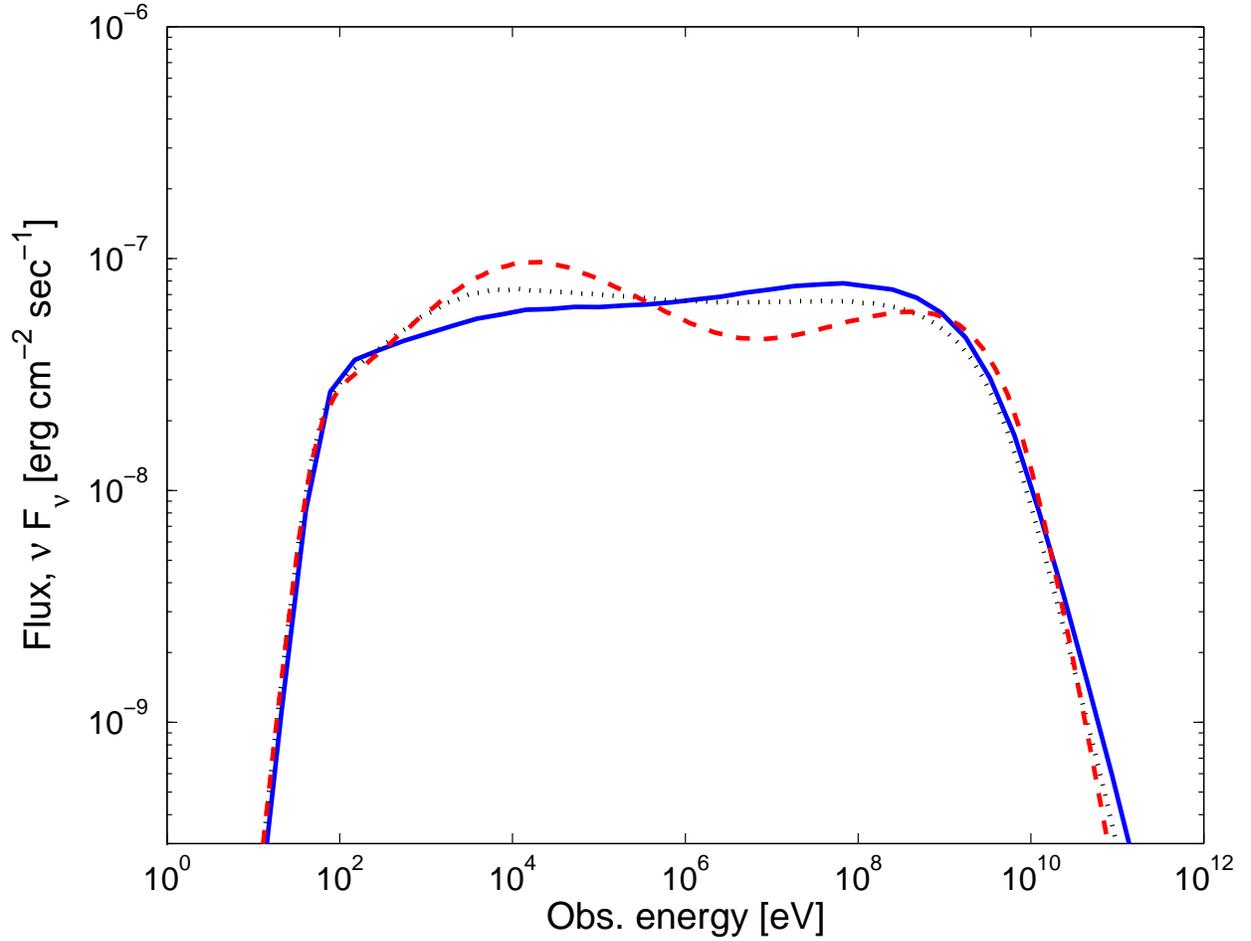}
\caption{Dependence of spectra on the power law index $p$ of accelerated 
electrons, low compactness case. Results are shown for
$\Delta t = 10^{-2} {\rm \ s}, \Gamma = 300$, $l'=2.5$, and
$p=2.0$ (solid), 2.5 (dotted), 3.0 (dashed). 
All other parameters are the same as in Figure \ref{fig:results1}.
Spectra depend only weakly on $p$.
}
\label{fig:p1}
\end{figure}

\begin{figure}
\plotone{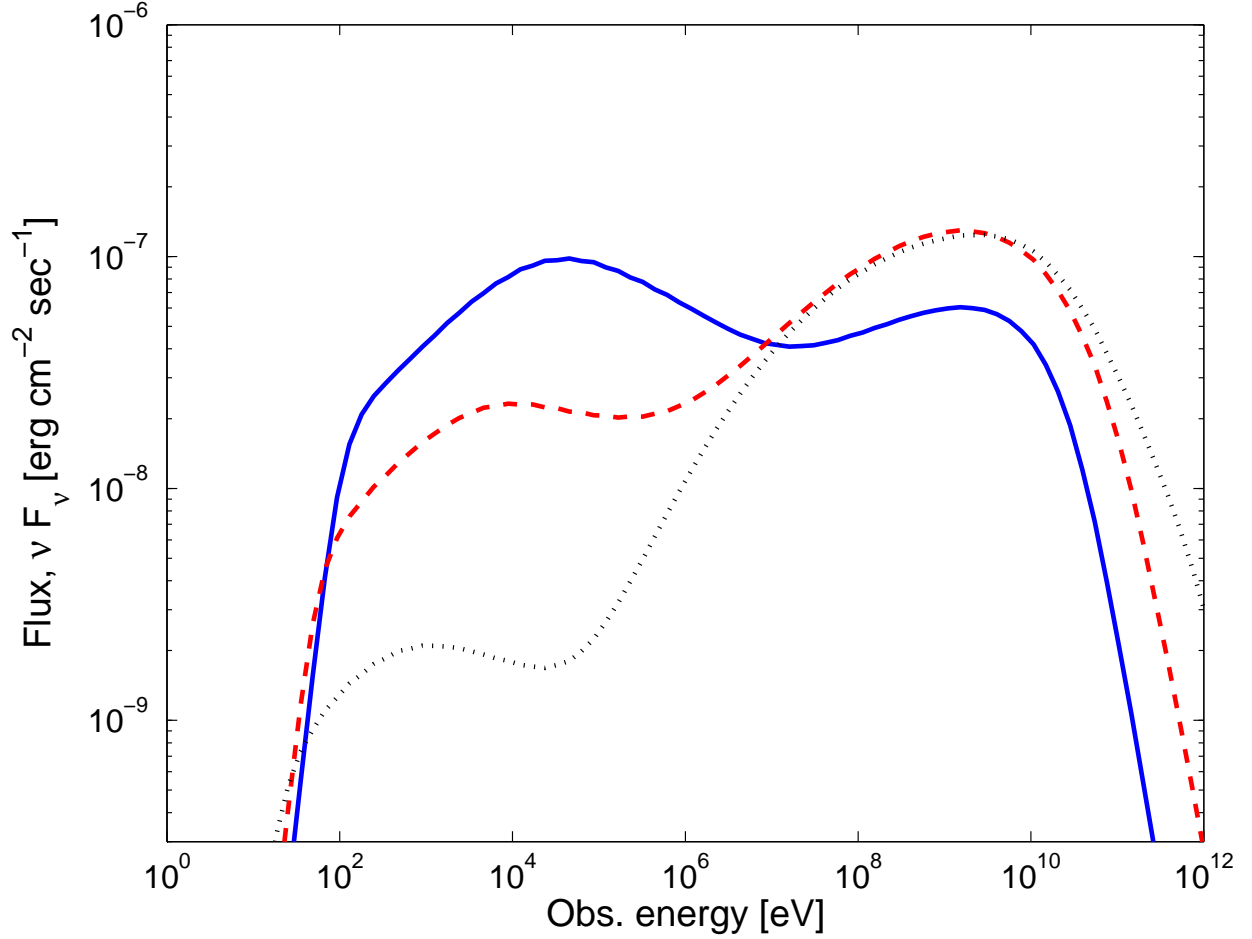}
\caption{Dependence on the fraction of thermal energy carried
by magnetic field, $\epsilon_B$, low compactness case. Results are shown for
$\Delta t = 10^{-3} {\rm \ s},  \Gamma = 600$, $l'=0.8$., and
$\epsilon_B = 0.33$ (solid), $\epsilon_B = 10^{-2}$ (dashed), 
$\epsilon_B = 10^{-4}$ (dotted). 
All other parameters are the same as in Figure \ref{fig:results1}. 
The ratio of fluxes at 1~\GeV and 0.1~MeV is a good indicator for the ratio
$\epsilon_B:\epsilon_e$.
}
\label{fig:xi_B}
\end{figure}

\begin{figure}
\plotone{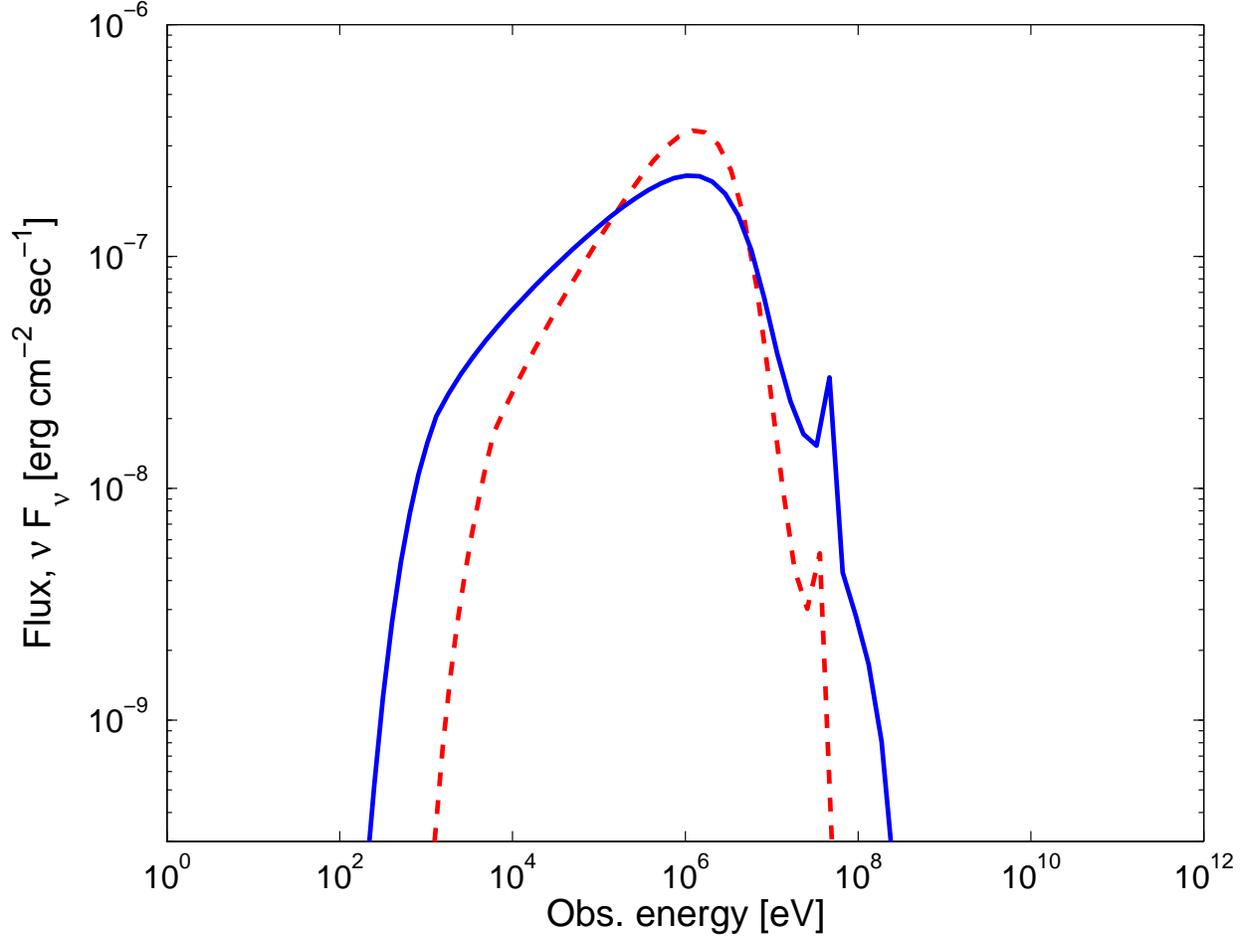}
\caption{
Time averaged spectra obtained for high compactness.
Results are shown for $L=10^{52}$~erg, $\epsilon_e=\epsilon_B=10^{-0.5}$, $p=3$ (all cases) and: $\Delta t = 10^{-4} {\rm \ s}$ , $\Gamma = 300$, (solid);
$\Delta t = 10^{-5} {\rm \ s}$ , $\Gamma = 300$.
The compactness parameter is $l' = 250, 2500$, respectively.
The scattering optical depth at the end of the dynamical time 
is 13, 56 respectively. The peaks observed at $\sim 80 \MeV$ result from pair annihilation.
Luminosity distance $d_L=2 \times 10^{28}$ and $z=1$ were assumed.
}
\label{fig:results2}
\end{figure}

\begin{figure}
\plotone{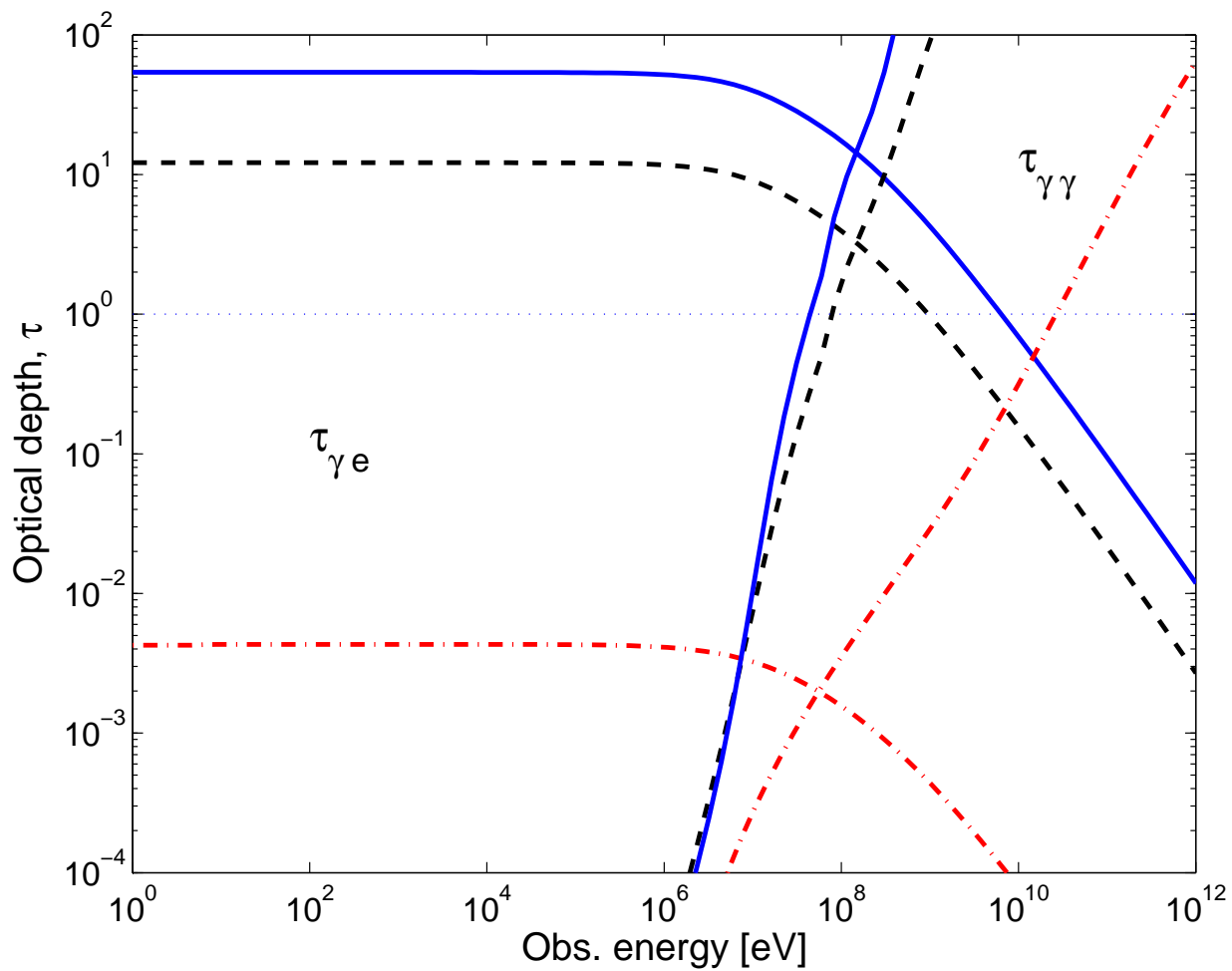}
\caption{Energy dependent optical depths for pair production and scattering.
Solid: $\Delta t = 10^{-5} {\rm \ s}, p=3.0$, $\Gamma = 300$, $l'=2500$;
Dashed: $\Delta t = 10^{-4} {\rm \ s}, p=3.0$, $\Gamma = 300$, $l'=250$;
Dash-dotted:  $\Delta t = 10^{-4} {\rm \ s}, p=3.0$, $\Gamma = 1000$, $l'=0.6$.
All other parameters are the same as in Figure \ref{fig:results1}.
}
\label{fig:tau}
\end{figure}

\begin{figure}
\plotone{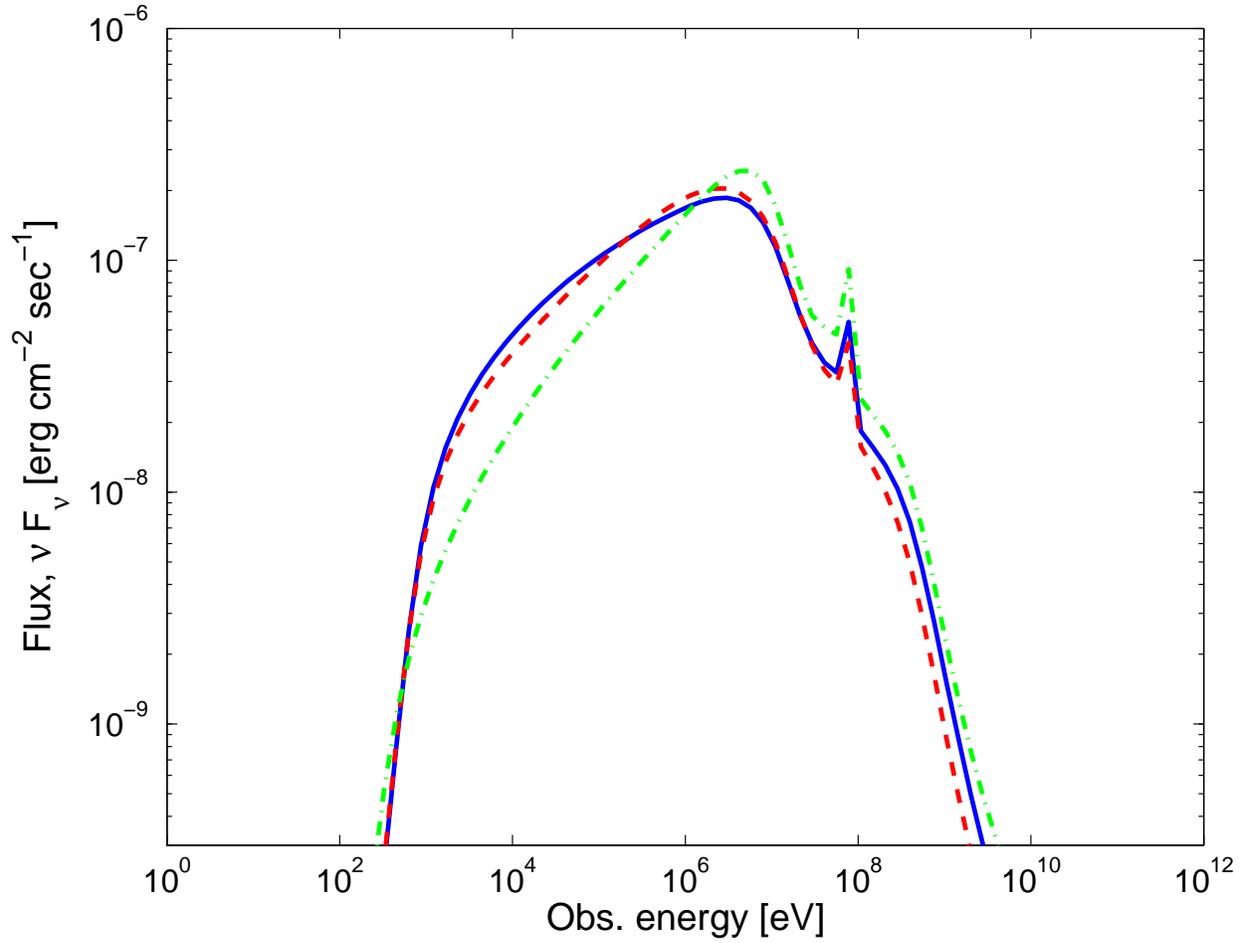}
\caption{Dependence of spectra on the power law index $p$ of accelerated 
electrons, and on the fraction of thermal energy carried
by magnetic field, $\epsilon_B$, for high compactness. Results shown for 
$\Delta t = 10^{-4} {\rm \ s}, \Gamma = 300$, $l'=250$ and 
$p=2.0$, $\epsilon_B = 0.33$ (solid); $p=3.0$, $\epsilon_B = 0.33$ (dashed);
$p=2.0$, $\epsilon_B = 0.01$ (dash-dotted).
All other parameters are the same as in Figure \ref{fig:results1}.
Spectra depend only weakly on $p$ and $\epsilon_B$.
}
\label{fig:p3}
\end{figure}

\begin{figure}
\plotone{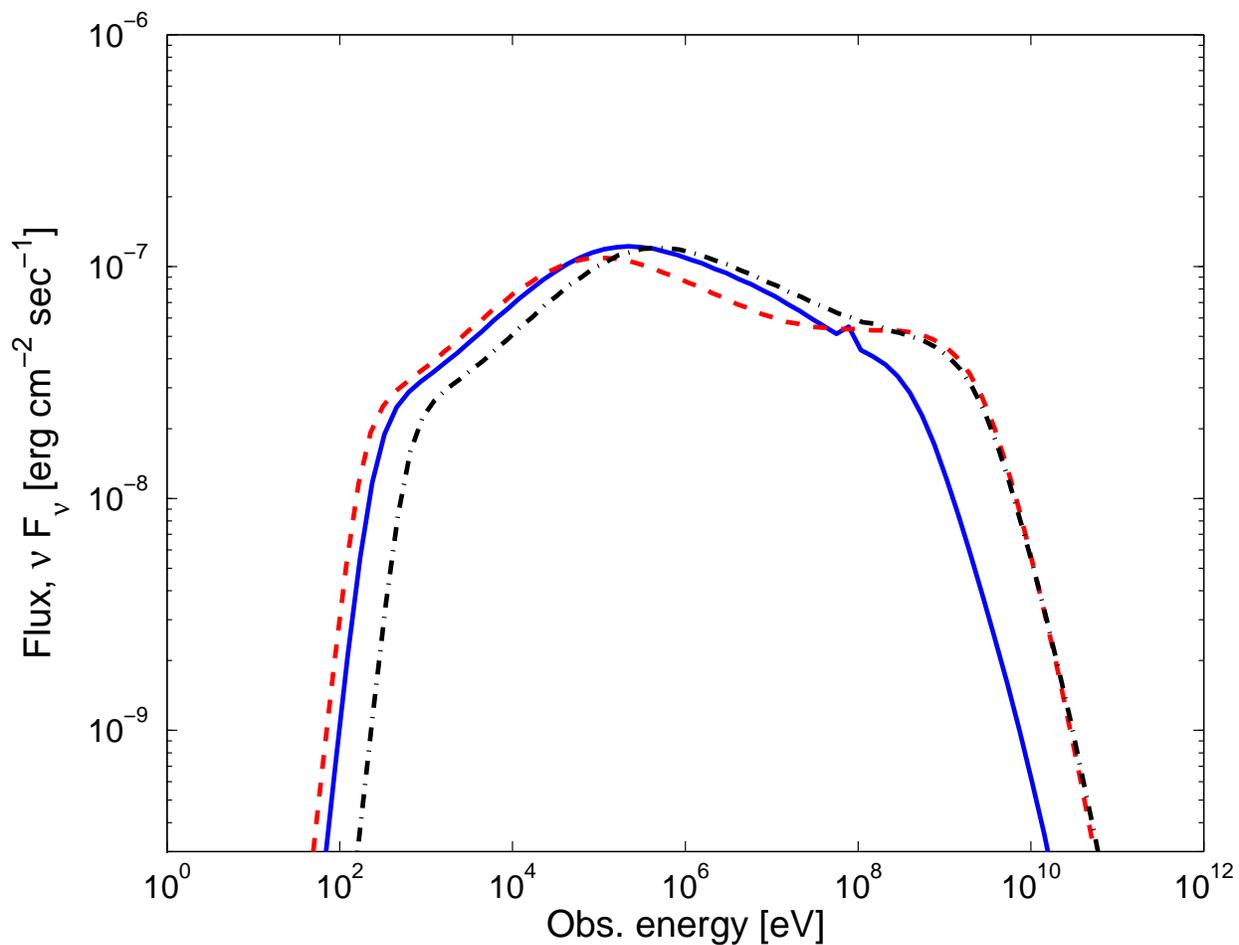}
\caption{Time averaged flux obtained for intermediate compactness.
Solid:  $\Delta t = 10^{-3} {\rm \ s}$ , $\Gamma = 300$;
Dashed: $\Delta t = 10^{-3} {\rm \ s}$ , $\Gamma = 400$;
Dash-dotted: $\Delta t = 10^{-4} {\rm \ s}$ , $\Gamma = 600$.
All the other fireball model parameters are the same as in figure 
 \ref{fig:results1}.
The compactness parameter is $l' = 25, 6, 8$, respectively.
}
\label{fig:results3}
\end{figure}

\begin{figure}
\plotone{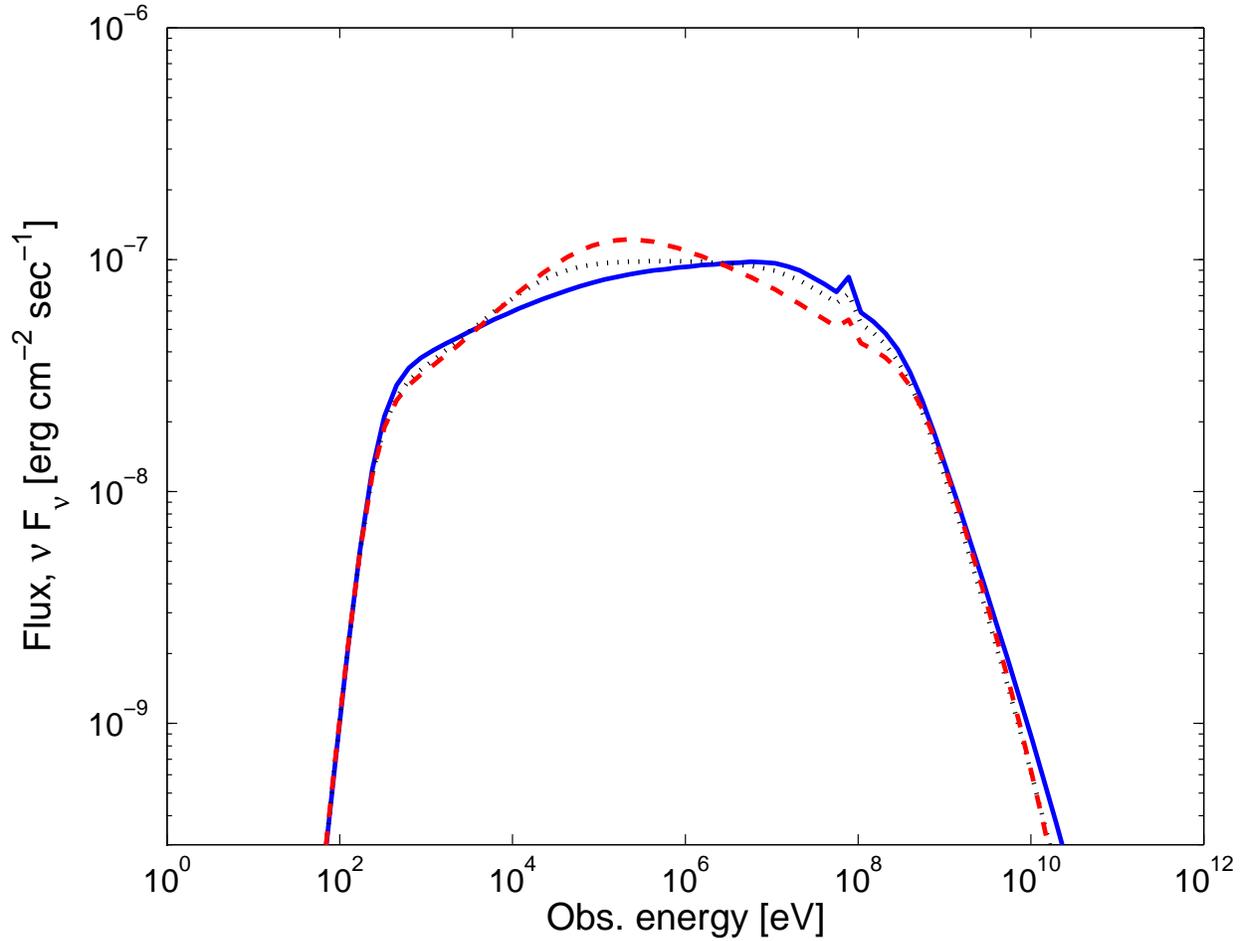}
\caption{Dependence of spectra on the power law index $p$ of accelerated 
electrons, intermediate compactness. Results shown for
$\Delta t = 10^{-3} {\rm \ s}, \Gamma = 300$, $l'=25$, and
$p=2.0$ (solid), 2.5 (dotted), 3.0 (dashed).
All other parameters are the same as in Figure \ref{fig:results1}.
}
\label{fig:p2}
\end{figure}

These conclusions are consistent with the results of our detailed numerical calculations. We have presented numeric results of calculations of prompt GRB spectra within the fireball model framework, using a time dependent numerical code which describes cyclo-synchrotron
emission and absorption, inverse and direct Compton scattering, and
$e^\pm$ pair production and annihilation. We have shown that the spectral shape depends mainly on the compactness parameter, which is most sensitive to the fireball Lorentz factor $\Gamma$, $l'\propto\Gamma^{-5}$. For large compactness (small $\Gamma$), $l'>100$, the spectra peak at $\sim1$~MeV, show steep slopes at lower energy, $\varepsilon^2 dN/d\varepsilon\propto \varepsilon^\alpha$ with $0.5<\alpha<1$, and show a sharp cutoff at $\sim10$~MeV (see fig.~\ref{fig:results2}). The spectra depend only weakly in this regime on the power-law index $p$ of accelerated electrons and on the magnetic field energy fraction $\epsilon_B$ (see fig.~\ref{fig:p3}).
For small to moderate compactness (large $\Gamma$), $l'\lesssim10$, spectra extend to $\gtrsim10$~GeV (see figures~\ref{fig:results1},~\ref{fig:results3}). The spectrum at lower energy depends only weakly on $p$ (see fig.~\ref{fig:p1}), but strongly on $\epsilon_B$ (see fig.~\ref{fig:xi_B}). For magnetic field close to equipartition, spectra peak at $\sim0.1$~MeV for $l'\sim10$ and extend to higher energy with spectral index $\alpha=10^{0\pm0.5}$.

\begin{figure}
\plotone{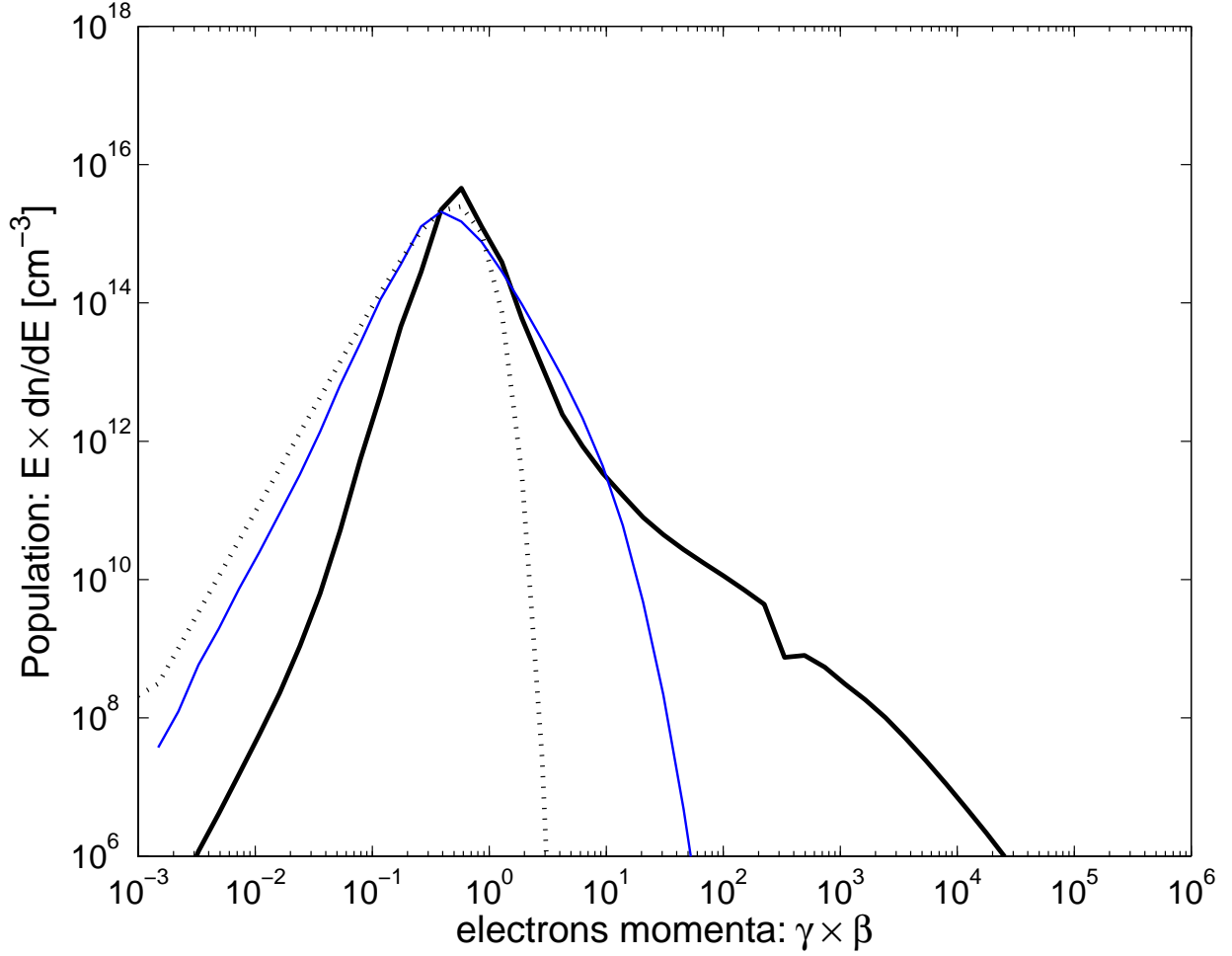}
\caption{Electrons energy distribution, before and after the adiabatic expansion
phase for
$\Delta t = 10^{-4} {\rm \ s}, p=3.0$, $\Gamma = 300$, $l'=250$
(and all other parameters the same as in Figure \ref{fig:results1}).
Thick: distribution at the end of the dynamical time;
Thin: distribution at the end of the adiabatic expansion.
The dotted line shows a Maxwellian distribution with temperature 
$\theta\equiv kT/m_ec^2 \approx 0.08$.
The particle distribution approaches Maxwellian only at the end 
of the adiabatic expansion phase.
}
\label{fig:elec_ad}
\end{figure}

Since moderate to large compactness is required for a synchrotron peak at $\sim1$~MeV, the effects of pair-production and direct and inverse-Compton scattering by pairs are predicted to be large for observed GRBs. In this case, simple analytic approximations of synchrotron-self-Compton emission do not provide an accurate description of the emergent spectra. In particular, Compton scattering by the pairs, which are accumulated at intermediate energy $\theta\equiv\varepsilon/m_e c^2 \approx 0.1$ (see figure~\ref{fig:elec_ad}), results in a steep slope, $0.5\le\alpha\le1$, in the 1~\keV to 1~\MeV band. Still steeper slope, $\alpha \approx 3$, is obtained at lower energies, below the self-absorption frequency determined by the quasi-thermal pair distribution, $\varepsilon_{ssa}^{ob.}\approx 10^{0.5\pm0.5} \keV$ (see Eq.~\ref{eq:eps_ssa_thermal}). These steep slopes may account for the steep slopes observed at early times in some GRBs \citep[see][]{Frontera,Preece,Ghirlanda}.

The spectra presented in this paper (\S~\ref{sec:results}) describe the emission resulting from a single collision within the expanding fireball wind, for various choices of model parameters ($L$, $\Delta t$, $\Gamma$ etc.). Observed spectra are expected to be combinations of spectra produced by many single-shell collisions, each characterized by different parameters. This is due to the fact that at any given time a distant observer is expected to receive  radiation from many collisions taking place at various locations within the wind. Moreover, observed spectra are inferred from measurements where the signal is integrated over time intervals longer than that expected for the duration of a single collision. A detailed comparison with observations requires a detailed model describing the distribution of single-shell collision parameters within the fireball wind \citep[see, e.g., ][]{Daigne98,Panaitescu99,GSW01}. The construction and investigation of such models are beyond the scope of this manuscript.

\begin{figure}
\plotone{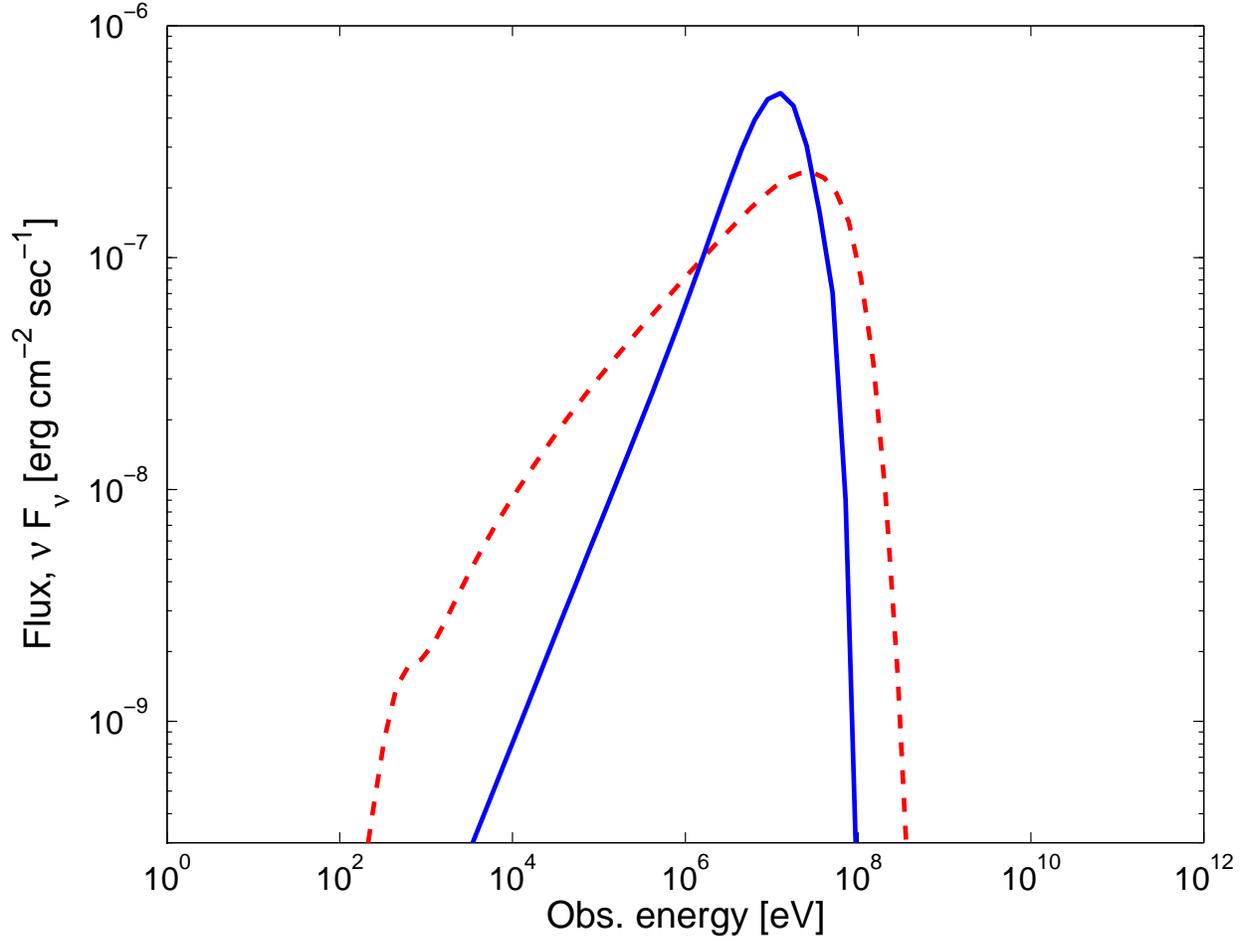}
\caption{Time averaged flux for the
quasi thermal Comptonization scenario.
Fireball model parameters assumed are 
$L=10^{52} {\rm \ erg \ s^{-1}}$, 
$\Gamma = 300$,
$\epsilon_e = \epsilon_B = 0.33$.
$\Delta t = 10^{-3} {\rm \ s}$ (dashed), 
$\Delta t = 10^{-4} {\rm \ s}$ (solid).
}
\label{fig:ghisellini}
\end{figure}

We have shown (\S~\ref{sec:ghis}, figure~\ref{fig:ghisellini}) that the quasi-thermal Comptonization scenario \citep[e.g. ][]{Ghisellini99}, in which kinetic energy dissipated in shocks is continuously distributed roughly equally among all electrons, leads to high peak energy, $\gtrsim 30 \MeV$. This scenario may not account therefore for observed GRB spectra.  

Clearly, the most stringent constraints on $l'$, and hence on $\Gamma$, will be provided by measurements of the spectra at high energy $\gtrsim0.1$~GeV. The fluxes predicted by the model are detectable by GLAST\footnote{http://www-glast.stanford.edu}.
For large compactness, where emission is strongly suppressed above 0.1~GeV, the model predicts a steep spectrum at low energy, which is weakly dependent on other model parameters (figure~\ref{fig:p3}). For small compactness, where strong emission is expected above 10~MeV, the low energy spectrum depends mainly on $\epsilon_B$ (fig.~\ref{fig:xi_B}). Comparison of the flux at $\sim1$~GeV and $\sim100$~keV 
(available, e.g.  with SWIFT\footnote{http://www.swift.psu.edu})
may therefore allow to determine the fireball magnetic field strength. Stringent constraints on $p$, the spectral index of the energy distribution of accelerated electrons, will be difficult to obtain, due to the weak dependence of spectra on this parameter.

\acknowledgements
This work was supported in part by a Minerva grant and by an ISF grant.

%\end{document}

%=========

%=========

\end{document}